\documentclass[bibyear]{aa} 

\usepackage{amsmath}
\usepackage{amsfonts}
\usepackage{amssymb}
\usepackage{graphicx}
\begin{document} 

\title{Faint AGNs at $z>4$ in the CANDELS GOODS-S field: looking for contributors to the reionization of the Universe}

   \author{E. Giallongo\inst{1}, A. Grazian\inst{1}, F. Fiore\inst{1}, A. Fontana\inst{1}, L. Pentericci\inst{1}, E. Vanzella\inst{2}, M. Dickinson\inst{3}, D. Kocevski\inst{4}, M. Castellano\inst{1}, S. Cristiani\inst{5}, H. Ferguson\inst{6}, S. Finkelstein\inst{7}, N. Grogin\inst{6},  N. Hathi\inst{8}, A. M. Koekemoer\inst{6}, J. A. Newman\inst{9}, M. Salvato\inst{10}
          }

   \institute{INAF-Osservatorio Astronomico di Roma, via Frascati 33, 00040 Monteporzio, Italy
\and
INAF-Osservatorio Astronomico di Bologna, via Ranzani 1, 40127, Bologna, Italy
\and
   NOAO, 950 N. Cherry Avenue, Tucson, AZ 85719, USA
\and
  Department of Physics and Astronomy, University of Kentucky, Lexington, KY 40506, USA
   \and
     INAF-Osservatorio Astronomico di Trieste, via G.B. Tiepolo 11, 34131, Trieste, Italy
      \and
        Space Telescope Science Institute, 3700 San Martin Drive, Baltimore, MD 21218, USA
         \and
           Department of Astronomy, The University of Texas at Austin, Austin, TX 78712, USA
            \and
            Aix Marseille Universit\'{e}, CNRS, LAM (Laboratoire d'Astrophysique de Marseille) UMR 7326, 13388, Marseille, France
            \and
  Department of Physics and Astronomy, University of Pittsburgh, 3941 O’Hara Street, Pittsburgh, PA 15260, USA
   \and
      Max Planck Institute for extraterrestrial Physics, Giessenbachstrasse 1, D-85748 Garching bei Munchen, Germany
         }
   \date{}
   \authorrunning{E. Giallongo et al.}
   \titlerunning{Faint AGNs at $z>4$ in the CANDELS GOODS-S field}
 
\abstract
{Establishing the number of faint AGNs at $z=4-6$ is crucial to understand their cosmological importance as main contributors to the reionization of the Universe. }
{In order to derive the AGN contribution to the cosmological ionizing emissivity we have selected faint AGN candidates at $z>4$ in the CANDELS GOODS-South field which is one of the deepest fields with extensive multiwavelength coverage from Chandra, HST, Spitzer and various groundbased telescopes.} 
{We have adopted a relatively novel criterion. As a first step high redshift galaxies are selected in the NIR $H$ band down to very faint levels ($H\leq27$) using reliable photometric redshifts. This corresponds at $z>4$ to a selection criterion  based on the galaxy rest-frame UV flux.  AGN candidates are then picked up from this parent sample if they show X-ray fluxes above a threshold of $F_X\sim 1.5\times 10^{-17}$ erg cm$^{-2}$ s$^{-1}$ (0.5-2 keV), correponding to a probability of spurious detections of $2\times 10^{-4}$ in the deep X-ray 4 Msec Chandra image.}
{We have found 22 AGN candidates at $z>4$ and we have derived the first estimate of the UV luminosity function in the redshift interval $4<z<6.5$ and absolute magnitude interval $-22.5\lesssim M_{1450} \lesssim -18.5$ typical of local Seyfert galaxies. The faint end of the derived luminosity function is about two/four magnitudes fainter at $z\sim 4-6$ than that derived from previous UV surveys.  We have then estimated ionizing emissivities and hydrogen photoionization rates in the same redshift interval under reasonable assumptions and after discussion of possible caveats, the most important being the large uncertainties involved in the estimate of photometric redshift  for sources with featureless, almost power-law SEDs and/or low average escape fraction of ionizing photons from the AGN host galaxies. Indeed both effects could in principle significantly reduce the estimated average volume densities and/or ionizing emissivities especially at the highest redshifts.}
{ We argue that, under reasonable evaluations of possible biases, the probed AGN population can produce at $z=4-6.5$ photoionization rates consistent with that required to keep highly ionized the intergalactic medium  observed in the Lyman-$\alpha$ forest of high redshift QSO spectra, providing an important contribution to the cosmic reionization.}

   \maketitle
%

\section{Introduction}
The process of cosmic reionization remains one of the most important and puzzling problems for cosmology.
The reionization of the universe, after its neutral phase followed by recombination, appears well completed at $z\sim 6$ as inferred from the high ionization level of the intergalactic medium (IGM) observed in the Lyman-$\alpha$ forest of high redshift QSO spectra (\cite{fan06}). The epoch when the process becomes important is still unknown although some hints come from the level of thompson scattering of the CMB radiation observed by WMAP and Planck and by the analysis of the kinematic Sunayev-Zeldovich effect (\cite{hinshaw13,george14}). The common wisdom is that in the epoch corresponding to the redshift interval $z\sim 6-10$ the Universe has quickly changed its thermal and ionization state. However the identification of the source populations emitting enough UV radiation to reionize the Universe is still an open issue.

In more than fifteen years extensive studies have been performed to identify the ionizing population among the high redshift star forming galaxies or the bright AGN population.
Since the apparent number density of bright QSOs and AGNs is rapidly decreasing at $z>3$ (e.g. \cite{masters12}) it is usually assumed that QSOs do not produce enough ionizing emissivity at $z>4$. Indeed estimates  based on simple parametric extrapolations both in luminosity and redshifts of the AGN luminosity function imply values for the cosmological emissivity $ 10^{23}$ ergs Mpc$^{-3}$ Hz$^{-1}$ s$^{-1}$ (e.g. \cite{haardt12}). This is one order of magnitude smaller than what required to keep the intergalactic medium highly ionized at  $z\sim 6$ despite the fact that each bright QSO is able to ionize its neighbourhood up to a distance of the order of e.g. 10 Mpc at $z\sim 5.5$ (\cite{prochaska09,songaila10,worseck14}).

As a consequence, the more common star forming galaxies are thought to be responsible for the cosmic reionization. Their contribution  depends on their abundance at low luminosities and on the average fraction of ionizing UV flux escaping from each galaxy into its neighbourhood.
At very high redshifts, $z>5$, direct measures of the ionizing Lyman continuum flux from galaxies are not viable due to the saturated hydrogen absorption by the IGM observed in QSO spectra. Estimates at lower $z\sim 3-4$ become feasible but the values so far appear discrepant relying on specific assumptions and interpretations. For example,
recent analyses suggest average ionizing escape fractions of $\sim 10-15$\% (\cite{nestor11,mostardi13}). But their $z\sim 3$ candidates show ionizing continuum often slightly shifted in position respect to the non-ionizing flux, casting serious doubt of significant contamination by foreground low redshift interlopers  (\cite{vanzella12}). Other estimates based on spectroscopic and very deep broad or narrow band imaging  gave only upper limits  $<5$\% (\cite{vanzella10,boutsia11}). At present the results on the estimated ionizing escape fraction for this relatively bright star forming population at $z\sim 3$ are not conclusive.

Faintest dwarf high redshift galaxies could behave in a different way showing larger escape fractions (see e.g. \cite{fontanot14}) and could be significant contributors to the cosmic reionization (\cite{ferrara13,wise14,yue14}). However some assumptions on the abundance at $z\sim 7-10$ of the faintest  galaxies down to $M_{UV}\sim -13$ and on their escape fractions ($\sim 15-80$\%) are needed to make the star forming galaxy population a major contributor to the reionization (e.g. \cite{finkelstein12}, \cite{robertson13}). A recent evaluation by \cite{finkelstein14} indicates escape fractions $\sim 13$\% to keep the IGM ionized up to $z\sim 6$ and midly ionized to $z\sim 8$ although uncertainties are very large.
Specific searches are starting in lensing galaxy clusters to measure the escape of Lyman continuum flux in intrinsically faint galaxies which are gravitationally lensed and amplified in luminosity by the intervening cluster (see e.g. \cite{vanzella12b,ishigaki14}). 

All these uncertainties in the galaxy contribution  leave open the possibility that AGNs could still play a leading role in contributing to the cosmological ionizing background although the estimate of their contribution is mainly affected by the poor knowledge of the faint end slope of their luminosity function especially at high redshifts due the lack of deep AGN surveys at various wavelengths (\cite{shankar07}).

The traditional view based on an evolutionary ionizing QSO background increasing from the local value out to $z=2-3$ and then quickly decreasing (see e.g \cite{haardt12}) is changing thanks to the recent multiwavelength deep surveys at $z>3$  (\cite{glikman11,civano11,fiore12}) showing the presence of a considerable number of previously unknown faint AGNs able to produce a rather steep luminosity function. The presence of this faint population is changing our estimate of the AGN contribution to the ionizing UV background although the selection of faint AGNs at the highest redshifts $z\sim 6$ is difficult with the current instrumentation. 
Most important, the inclusion of X-ray detection in the selection methods enables to extend the knowledge of the luminosity function to even fainter X-ray limits (\cite{civano11,fiore12}) of the order of L(2-10 KeV) $> 10^{43}$ erg s$^{-1}$ at $z> 4$. In particular, the novel technique proposed by Fiore et al. (2012) relies on the selection of faint high redshift AGNs among the high $z$ galaxies selected in NIR images (e.g. H band) which show any detection in deep and high resolution, Chandra X-ray images. 

A first estimate of the volume density of faint AGNs could also have important implications concerning the abundance and mass of the supermassive black hole seeds
(\cite{volonteri12}) and their early growth (\cite{volonteri14}). Important constraints are also expected for the initial spectrum of the density perturbations related to the nature, cold or warm, of the dark matter (see e.g. \cite{menci13}).

In \cite{giallongo12} we made a first attempt to estimate the possible contribution to the cosmic reionization by faint high $z$ AGNs based on the results provided by deep X-ray surveys and on a joint galaxy-AGN model prediction. We found that faint AGNs could be able to reionize the high redshift IGM provided that their average volume density only gradually decline from $z\sim 3$ to $z\sim 6.5$ as predicted by the model.

In the present paper we have produced a sample of galaxies  selected in the Candels/GOODS-S/ChandraDeepField-South field using one among the deepest NIR, optical and X-ray database and we have built a sample of AGN candidates based on X-ray detection in the same H band galaxy position, following the recipe as in \cite{fiore12}. In Sect. 2 we describe the CANDELS photometric parent catalog of high redshift galaxies. In Sect 3. we describe the AGN selection by means of the Chandra 4Ms image. In Sect. 4 we present an estimate of the faint end AGN UV luminosity function. In Sect. 5 we show predictions about the AGN ionizing emissivity and photoionization rate and finally in Sect. 6 we discuss some possible caveats.
Throughout the paper we adopt round cosmological parameters  $\Omega_{\Lambda}=0.7$, $\Omega_{0}=0.3$, and Hubble constant $h=0.7$ in units of 100 km/s/Mpc.
Apparent magnitudes are in the AB photometric system.

\section{The CANDELS GOODS-S catalog}

The CANDELS area (\cite{grogin11}, \cite{koekemoer11})  with deep Chandra coverage is providing the optimal dataset in terms of deep NIR photometry and relatively wide area. Indeed the NIR images obtained with the Wide Field Camera 3 (WFC3) are particular important in terms of angular resolution and depth, sampling the AGN UV emission ($<3000$ {\AA}) at $z>4$. Among the various CANDELS fields we have selected in this preliminary analysis the GOODS-South field where the deepest 4Ms Chandra images are optimal  for the selection of high redshift AGNs. The field covers an area of $\sim$170 sq. arcmin at a mean depth of H=27.5. The dataset also includes very deep imaging by the IRAC instrument on board the {\sl Spitzer Space Telescope} coming from the {\sl Spitzer Extended Deep Survey} (SEDS; \cite{ashby13}) and  covering the CANDELS fields to 26 AB mag ($3\sigma$) at both 3.6 and 4.5 $\mu$m. The Spitzer infrared bands sample the rest-frame optical region of AGNs at $z>4$. We adopted the official photometric catalog released by \cite{guo13} selecting galaxies in the WFC3 H band. The total number of sources detected in the GOODS-South fields at a mean depth of H=27.5 is about 28500.

\begin{table*}
\caption{Area and magnitude limits of the CANDELS GOODS-South and HUDF
fields}
\label{table:magcompl}
\centering
\begin{tabular}{c | c c  r r}
\hline\hline
Field & Area & $H_{160}$ Mag. compl.  & $N_{gal}$ & $N^{high-z}_{gal}$ \\
 & $arcmin^2$ & 50\% &  $H\leq 27$ &  $4\leq z\leq 6.5$ \\
\hline
GOODS-South \#1               &  11.05 &  25.75 &     801 &   24 \\
GOODS-South \#2               &  25.03 &  26.00 &   2794 &   77 \\
GOODS-South \#3               &  50.47 &  26.50 &   7984 & 240 \\
GOODS-South \#4               &  77.18 &  27.00 & 15310 & 698 \\
GOODS-South \#5 (HUDF)   &    5.18 &  27.75 &   1672 &   74 \\
\hline
TOTAL                                 & 168.91 &     -     & 26963 & 1113 \\
\hline
\end{tabular}
\end{table*}

Due to the complexity of the GOODS-South exposure map in the H band, the field is composed by several sub-areas with the associated magnitude limits. Simulations have been performed to assess the detection completeness as described in \cite{grazian14}.  In the shallowest area, the incompleteness corrections as a function of magnitude have been estimated  comparing the observed galaxy number counts with that in the deepest HUDF area. For the other areas of intermediate depths we scaled the correction factors to fainter magnitudes 
taking into account the increasing s/n ratio. Table 1 provides the magnitude corresponding to a 50\% level of completeness in each of the areas covered by different sub-regions. The numbers of galaxies at $H<27$ are also shown in each region and at different redshifts.

Several catalogs of photometric redshifts have been provided for the CANDELS GOODS-S field. \cite{dahlen13} have made a comparison analysis among the various codes.
Different galaxy templates have been used so that  the codes were sufficiently different to provide independent redshift estimates. They
have finally combined independent estimates obtained by using nine different codes.
Several spectroscopic redshifts available in the observed field were compared to the resulting photometric redshifts of the same objects to derive the optimal solution for the overall sample. The resulting average uncertainty is $|\Delta z|/(1+zspec)\sim 0.03$ at relatively bright magnitudes ($H_{160}\le 24$),
which increases to $\sim 0.06$ at $H_{160}\sim 26$ (\cite{dahlen13}). The fraction of outliers (defined as differences $|\Delta
z|/(1+zspec)> 0.15$) is confined to $\sim 3$\% and in particular is  low at $z>3.7$ due to the Lyman continuum break signal. A Bayesian method has been developed to provide redshift probability distributions (PDFs) that can be useful for error estimates. The PDFs of the AGN candidates of the present sample are shown in the Appendix.

In total, there are 1113 galaxies in the GOODS-South field with a robust spectroscopic or photometric redshift at $z\geq 4$ which are analysed for the search of any AGN activity.
\begin{table*}
\caption{Photometric and spectroscopic properties of the 22 AGN candidates}
\begin{center}
\begin{tabular}{c c c l l c c c c c cc}
\hline\hline
ID & RA & Dec & zphot & zspec & C & H & $M_{1450}$ & $\log F_X$  & $\log L_X$ & A& Previous Catalogs\\
 &  &  &  &  &  & &  & erg/cm$^2$/s  & erg/s & \\
\hline
273  & 53.1220463 & -27.9387409 & 4.49  & 4.762$^1$ & c & 23.96 & -21.37 & -15.97 & 43.80 & \#2 & M208,X403\\
4285 & 53.1664941 & -27.8716803 & 4.28 & & cf & 25.57  & -20.22 & -16.46 & 42.90 & \#3 & -- \\
4356 & 53.1465968 & -27.8709872 & 4.70 &  & cf & 26.36  & -18.44 & -16.38 & 43.40 & \#4 & M70437,L306,X485\\
4952 & 53.1605007 & -27.8649890 & 4.32 &  & c & 25.47  & -20.20 & -16.50 & 42.90 & \#3 & -- \\
5375 & 53.1026292 & -27.8606307 & 4.41 &  & c & 25.16  & -20.16 & -16.66 & 42.75 & \#4 & X331 \\
5501 & 53.1280240 & -27.8593930 & 5.39 &  & c & 25.71  & -20.23 & -16.45 & 43.10 & \#4 & -- \\
8687 & 53.0868634 & -27.8295859 & 4.23 &  & c & 26.90  & -19.19 & -16.43 & 42.90 & \#4 & -- \\
8884 & 53.1970699 & -27.8278566 & 4.52 &  & c & 25.74  & -19.04 & -16.77 & 42.65 & \#4 & -- \\
9713 & 53.1715890 & -27.8208052 & 5.86 & 5.70$^2$ & c & 26.54 & -19.87 & -16.46 & 43.15 & \#4 & HUDF322 \\
9945 & 53.1619508 & -27.8190897 & 4.34 & 4.497$^3$ & cd & 24.99 & -20.93 & -16.65 & 42.75 & \#4 & -- \\
11287 & 53.0689924 & -27.8071692 & 4.94 &  & c & 25.06 & -20.48 & -16.42 & 43.10 & \#4 & M8728\\
12130 & 53.1514304 & -27.7997601 & 4.43 & 4.62$^4$ & c & 25.54 & -20.60 & -16.58 & 42.85 & \#5 & HUDF3094 \\
14800 & 53.0211735 & -27.7823645 & 4.92 & 4.823$^5$ & c & 23.43 & -22.51 & -16.38 & 43.10 & \#3 &  M10548\\
16822 & 53.1115637 & -27.7677714 & 4.52 &  & c & 25.67 & -18.97 & -15.91 & 43.85 & \#4 & M70168,L245,X371\\
19713 & 53.1198898 & -27.7430349 & 4.84 &  & c & 25.31 & -18.14 & -16.48 & 43.00 & \#4 & E1516,X392\\
20765 & 53.1583449 & -27.7334854 & 5.23 &  & f & 24.44 & -21.06 & -16.29 & 43.25 & \#3 & E2551\\
23757 & 53.2036444 & -27.7143907 & 4.13 &  & c & 24.56 & -20.72 & -16.49 & 42.85 & \# 1 & -- \\
28476 & 53.0646867 & -27.8625539 & 6.26 &  & f & 26.77 & -19.03 & -16.60 & 43.10 & \#4 & M70407\\
29323 & 53.0409764 & -27.8376619 & 9.73 &  & cf & 26.33 & -19.50 & -15.96 & 44.00 & \#3 & M70340,L103,X156\\
31334 & 53.2131871 & -27.7816486 & 4.73 &  & f & 26.41 & -19.60 & -15.69 & 43.75 & \#4 & -- \\
33073 & 53.0547529 & -27.7368325 & 4.98 &  & c & 26.89 & -19.19 & -16.44 & 43.10 & \#2 & E2199\\
33160 & 53.0062504 & -27.7340678 & 6.06 &  & cf & 25.90 & -19.62 & -16.26 & 43.65 & \#3 & E2498,L57,X85\\
\hline
\end{tabular}
\end{center}

ID from \cite{guo13} catalog\\
Photometric redshifts (zphot) from \cite{dahlen13} catalog\\
Spectroscopic redshifts after: $^1$\cite{vanzella08}, $^2$\cite{hathi08}, $^3$M. Dickinson \& D. Stern private communication, $^4$\cite{rhoads09}, $^5$\cite{balestra10}.\\
Column C indicates the compactness of the sources. c=compact, d=diffuse, f=faint object\\
X-ray fluxes are computed in the 0.5-2 keV band and X-ray luminosities in the 2-10 keV band
M,E in Previous Catalogs are from \cite{fiore12}, L from \cite{luo08}, X from \cite{xue11}. Column A indicates the GOODS-S field area according to table 1
\end{table*}
It is to note that the photometric redshifts have been estimated using theoretical spectral energy distributions derived from stellar populations only. No contribution from the non stellar light typical of the AGN population has been included. Nevertheless, since the selection of faint AGNs is confined to $z\geq 4$, the redshift estimate is based on the dropout of the SED due to the neutral HI absorption by the IGM intervening along the line of sight, a feature independent of the main intrinsic spectral properties of the sources like e.g. continuum spectral shape or  escape fraction of ionizing photons from the AGN host galaxy. Indeed a large UV dropout is expected in any case at $z>4$ by optically thick intervening absorption by the IGM even if the ionizing escape fraction from the AGN host galaxy were high. Recently \cite{hsu14} have provided a catalog of photometric redshifts in the CANDELS GOODS-S area optimized for relatively bright X-ray detected AGNs. They add to the CANDELS photometric catalog new optical intermediate-band filters and explore a mixing of galaxy and AGN templates. Their recipe appears to provide good results in terms of redshift accuracy and low fraction of outliers when comparing with the spectroscopic redshift sub-sample. This is in particular true for bright sources with $H<23$. However for our much fainter sample having typically $H=24-26$ the use of shallower intermediate band filters and mixed AGN/galaxy templates does not improve the accuracy of photometric redshifts and the fraction of outliers.

\begin{figure*}
\centering
\scalebox{0.4}[0.4]{\rotatebox{0}{\hspace{0cm}\includegraphics{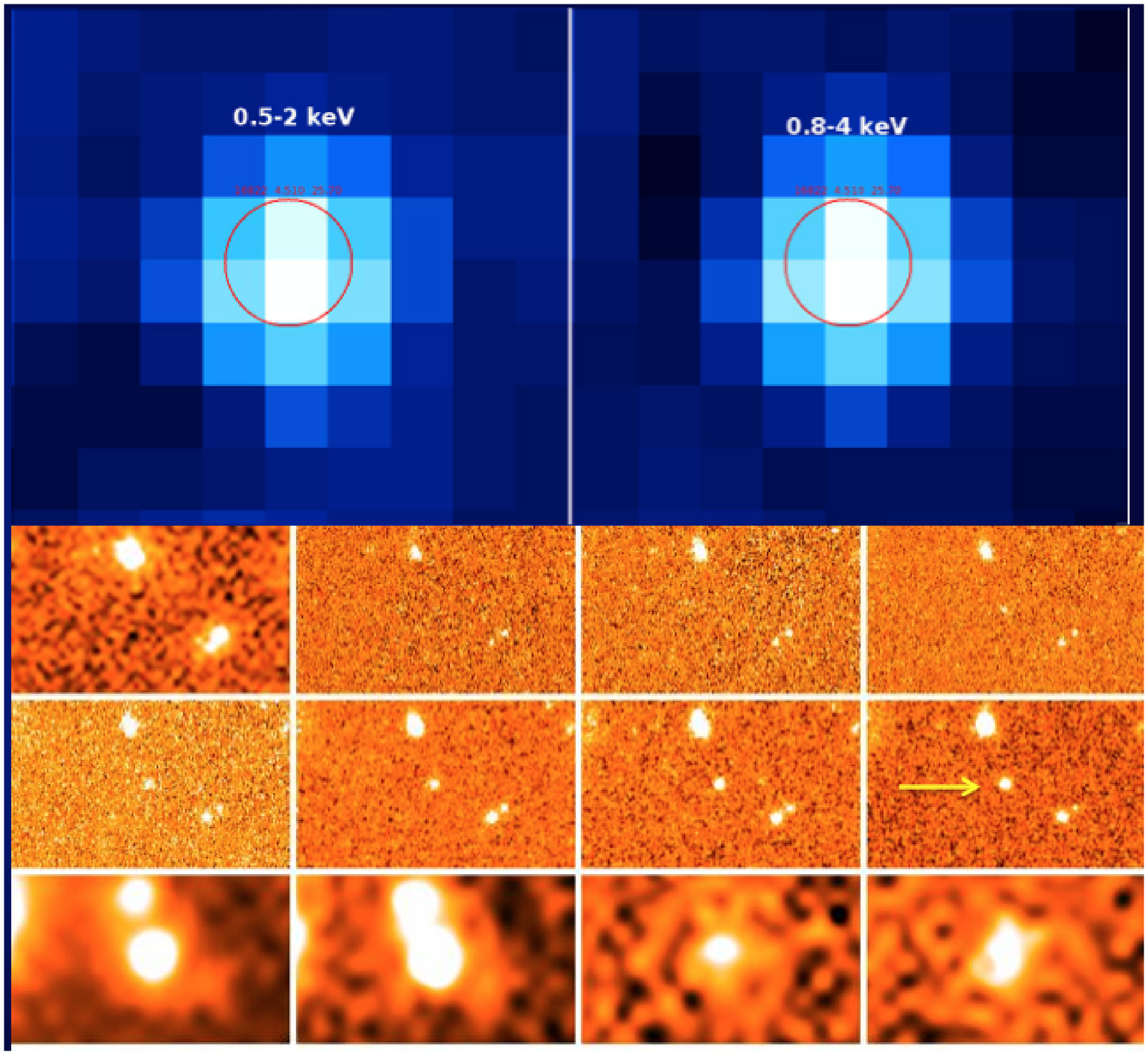}}}\scalebox{0.365}[0.365]{\hspace{0.5cm}\includegraphics{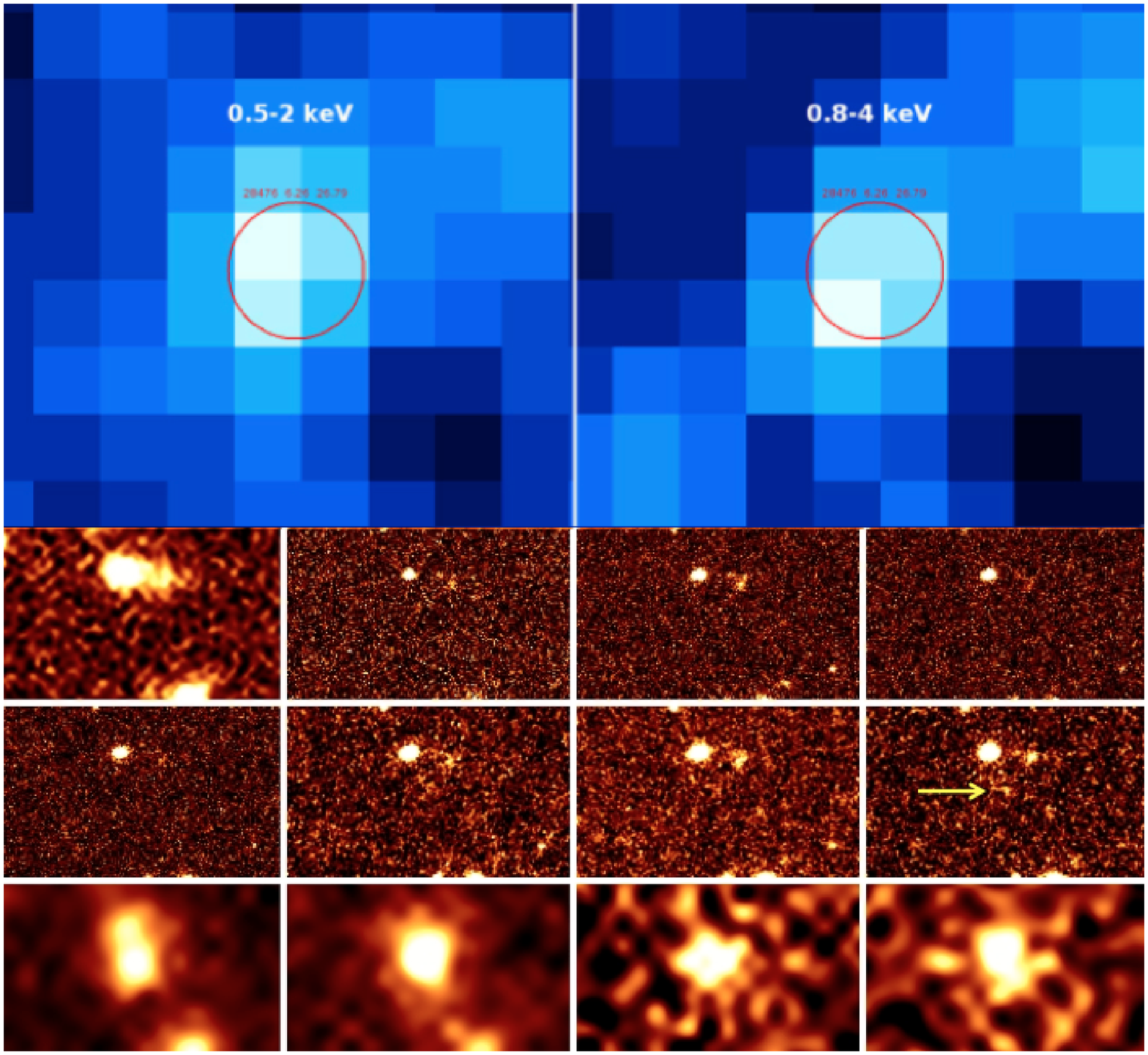}}
\caption{Image thumbs of two AGN candidates. From left to right the filters are X$_{0.5-2}$, X$_{0.8-4}$, U, B$_{435}$,V$_{606}$, I$_{775}$, z$_{850}$,  Y$_{1.05}$,  J$_{1.25}$,   H$_{1.6}$, IRAC $_{3.6}$, $_{4.5}$, $_{5.8}$, $_8$. Left box: object 16822 at $z=4.5$. Right box: object 28476 at $z=6.26$. The size of the optical/IR thumbs is $9\times 6$ arcsec$^2$. Red circles in the X-ray images show the expected position from the H band detection with a circular size of 1 arcsec}
\end{figure*}

\section{The AGN selection}

The selection of AGN candidates relies on the selection of galaxies at $z>4$ having $H<27$ from the general CANDELS catalog as described in the previous section. This corresponds to select sources on the basis of their detected rest-frame UV luminosity at $\lambda \sim 2000-3000$ {\AA}.
Following the procedure proposed in \cite{fiore12} we searched for any sign of AGN activity just measuring the X-ray flux in the expected H band position of the CANDELS sources having photometric or spectroscopic redshifts $z\geq 4$. In the following we briefly summarize the main steps.
Since X-ray data are structured in event files, preserving time, spectral and spatial information, a multidimensional source detection technique has been developed by \cite{fiore12} using spatial, spectral, and timing information. In particular, clustering of X-ray events in energy, time and spatial coordinates was investigated to efficiently detect X-ray bursty sources and X-ray sources characterized by specific features like strong lines or sharp edges, by reducing the elapsed time and/or the energy interval (i.e. the background) where the source detection is performed. Spectra were added at each galaxy position in different apertures from three to nine pixels. Sources are selected having more than 12 background subtracted counts in the wide 0.3-4 keV band. The background is evaluated from a map obtained from the original image after the exclusion of all bright sources. The background for each source was computed by normalizing an average background at the source off-axis to the source spectrum integrated in the 7-11 keV energy band, where the source contribution is negligible due to the decrease in the Chandra effective area. Source extraction aperture and contiguous energy band have been selected to minimize the Poisson probability for background fluctuation and to maximize the signal-to-noise ratio.
Extensive simulated data have been performed using about $10^5$ background spectra at the positions of the detected sources using the same exposure time, PSF etc.
studying the number of spurious detections as a function of threshold, aperture and energy band width. A combination of parameters has been adopted to keep the number of spurious detections smaller than one over 5000 spectra corresponding to a probability of spurious detection of $2\times 10^{-4}$. This implies that for the 1113 galaxies at $z>4$ we expect only 0.2 spurious AGN candidates in the field. Finally for each candidate position in the Chandra image we have checked that  the accuracy of the relative astrometry is $\lesssim 1$ arcsec (see the Appendix for an example).

We have detected 22 reliable AGN candidates with $z>4$ whose positions, redshifts, X-ray (0.5-2 keV) fluxes and H magnitudes  are shown in Table 2. Eleven of the detected sources were already in \cite{fiore12} catalog although with somewhat different photometric redshifts, four
in the GOODS-S catalogs of X-ray selected sources by \cite{luo08} and seven are in \cite{xue11} (four in common with \cite{luo08}).  Five redshifts have been confirmed spectroscopically either by the presence of strong Lyman-$\alpha$ intervening absorption trough (for the 9713 and 12130 objects observed in slitless low resolution spectroscopy with ACS camera at HST) or by the detection of Lyman-$\alpha$ emission. The detection of CIV at $z>4$ is usually not confirmed because out of the observed spectral region or expected in regions dominated by strong sky emission lines. The morphological appearance is indicated in column C where most candidates after visual inspection appear compact in H band with FWHM$\sim 0.2-0.4$ arcsec. Only one (9945) has a complex morphology. The remaining are too faint for any classification. It is to note the larger number of AGN candidates emerging from the combined H band selection plus X-ray detection respect to the direct X-ray selection. X-ray luminosities have been computed for the brightest sources adopting a standard single power-law with fixed slope $1.8$ of the power-law photon index and fitting a gas absorption. We derived hydrogen column densities of $\log N_H\sim 23.5$ for 273, 4356, 16822, 29323, 33160 and $\log N_H\sim 22.5$ for 33073, 20765. For the remaining X-ray
fainter objects we adopted $\log N_H\sim 22$. The reliability of the estimated X-ray absorption derived for the estimate of X-ray fluxes and luminosities in $z>4$ faint sources is affected by the higher rest-frame energy sampled, by the adopted slope in the power-law fitting and by other physical assumptions like covering factor and metallicity of the absorbing gas, as discussed in the last section. 
The derived X-ray luminosities in table 2 have an average value of $\log L_X\sim 43.2$ erg s$^{-1}$ and are more typical of Seyfert-like sources rather than of starburst galaxies although we can not exclude a significant contribution from stars to the X-ray luminosity of some peculiar source in our sample.
Analysis by \cite{fiore12}  of the brightest sources at $z\sim 3$ suggests that the possible presence of significant X-ray absorption is uncorrelated with the presence of Lyman-$\alpha$ and CIV emission lines and low dust extinctions as in the case where absorption is in general close to the emitting ionized region and inside the sublimation radius.

\begin{figure*}
\centering
\scalebox{0.9}[0.9]{\includegraphics[angle=-90,width=\hsize]{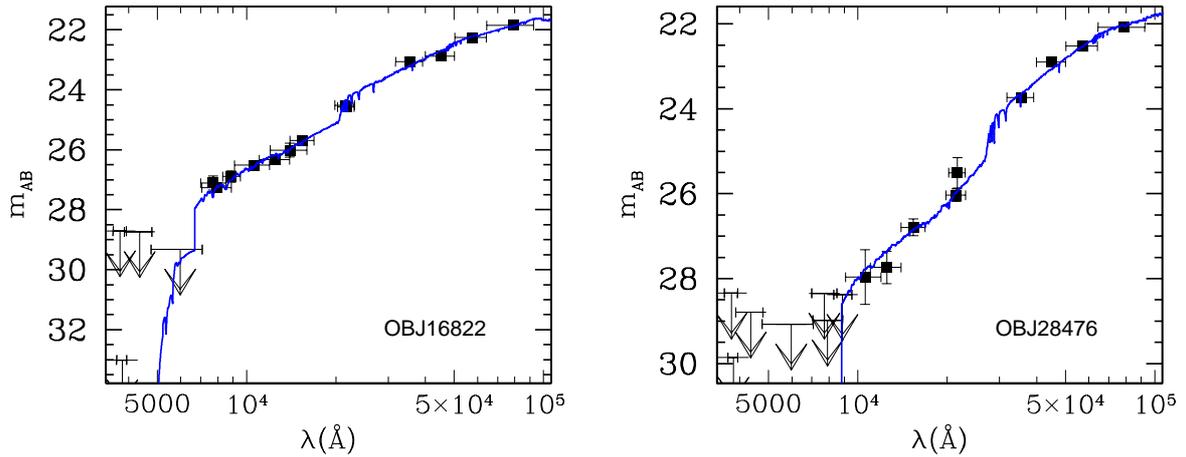}}
\caption{Spectral energy distributions of the objects as in figure 1}
\end{figure*}

Figure 1 shows as an example the thumbs of 2 sources with decreasing wavelength from the IR Spitzer images to the HST images as well as the X-ray image. The first source on the left, 16822, has a good signal-to-noise ratio and appears compact in the H band. The second AGN candidate  on the right, 28476,  is among the highest redshift sources in our sample. It is much fainter in the H band lacking any morphological information.

\begin{figure*}
\centering
\scalebox{0.5}[0.5]{\includegraphics[width=\hsize]{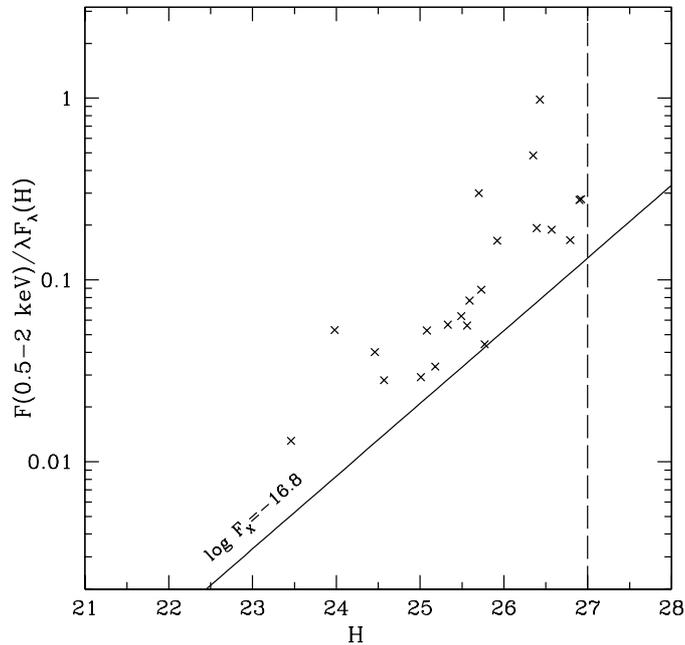}}
\begin{center}
\caption{X/H flux ratio as a function of H magnitude for the CANDELS GOODS-S sample. The X-ray flux limit adopted in the Chandra image is also shown together with the H magnitude limit $H=27$.}
\end{center}
\end{figure*}

Figure 2 shows the spectral energy distributions of the two candidates shown in figure 1 together with their galaxy spectral fits and the derived photometric redshifts. It is to note how faint are the selected AGNs  in the optical bands. The SEDs of all the selected AGN candidates are shown in the Appendix. This outlines the importance of the joint H band, X-ray detections to find out faint AGNs in multiwavelength surveys (\cite{fiore12}).
Figure 3 shows the X-ray over H-band flux ratio as a function of the H band magnitude for our CANDELS GOODS-S sample of AGN candidates. The straight line in the figure represents the locus of constant X-ray flux corresponding to the average detection threshold. Looking at the X/H ratio distribution derived at different H magnitudes it appears that only faint NIR $z\geq 4$ candidates which are relatively bright in the X-ray band can be detected in the Chandra 4Ms field.  In this respect our sample is biased against relatively X-ray faint AGNs, a feature that has to be taken into account in estimating the faint end of the AGN UV luminosity function as computed in the next section.
We note again that our sample is coming from a well defined magnitude selected galaxy sample in the UV rest-frame band. The X-ray detection is only used in this context to reveal the presence of AGN activity.

\section{AGN UV luminosity functions}

The estimate of the faint end of the AGN luminosity function in the UV rest frame has been performed with an extended version of the standard $1/V_{max}$ algorithm (\cite{schmidt68}) where different regions in the GOODS-S field with different magnitude limits were combined together in the volume estimate of each object (e.g. \cite{avni80}). Indeed, for each object and for each j-th region shown in table 1, an effective volume $V_{max}( j)$ is computed. For a given redshift interval $(z_{low}, z_{up})$, these volumes are confined between $z_{low}$ and $z_{lim} ( j)$, the latter being defined as the minimum between $z_{up}$ and the maximum redshift at which the object could have been observed within the magnitude limit of the $j-th$ region. The galaxy number density $\phi(M, z)$ in a given $(\Delta z, \Delta M)$ bin can be computed as follows:
\begin{equation}
\phi(M,z)=\frac{1}{\Delta M} \sum_{i=1}^n\left[ \sum _j \omega (j)  \int _{z_{low}} ^{z_{lim}(i,j)} \frac{dV}{dz} dz  \right]^{-1} 
\end{equation}
where $\omega( j)$ is the area in units of steradians corresponding to the region $j$ shown in table 1, $ n$ is the number of objects in the bin and $dV/dz$ is the comoving volume element per steradian (see e.g. \cite{salimbeni08}).

Rest frame UV 1450 {\AA} absolute magnitudes $M_{1450}$ were computed from apparent magnitudes in the filters closer to this rest frame wavelength for each redshift. For example, for AGNs at $z\sim 4.5$ the UV rest-frame absolute magnitudes have been directly derived from the observed HST $i_{775}$ magnitudes.
The absolute magnitudes $M_{1450}$ used for the evaluation of the luminosity function are shown for each object in Table 2.

\begin{figure*}
\centering
\scalebox{0.8}[0.8]{\includegraphics[width=\hsize]{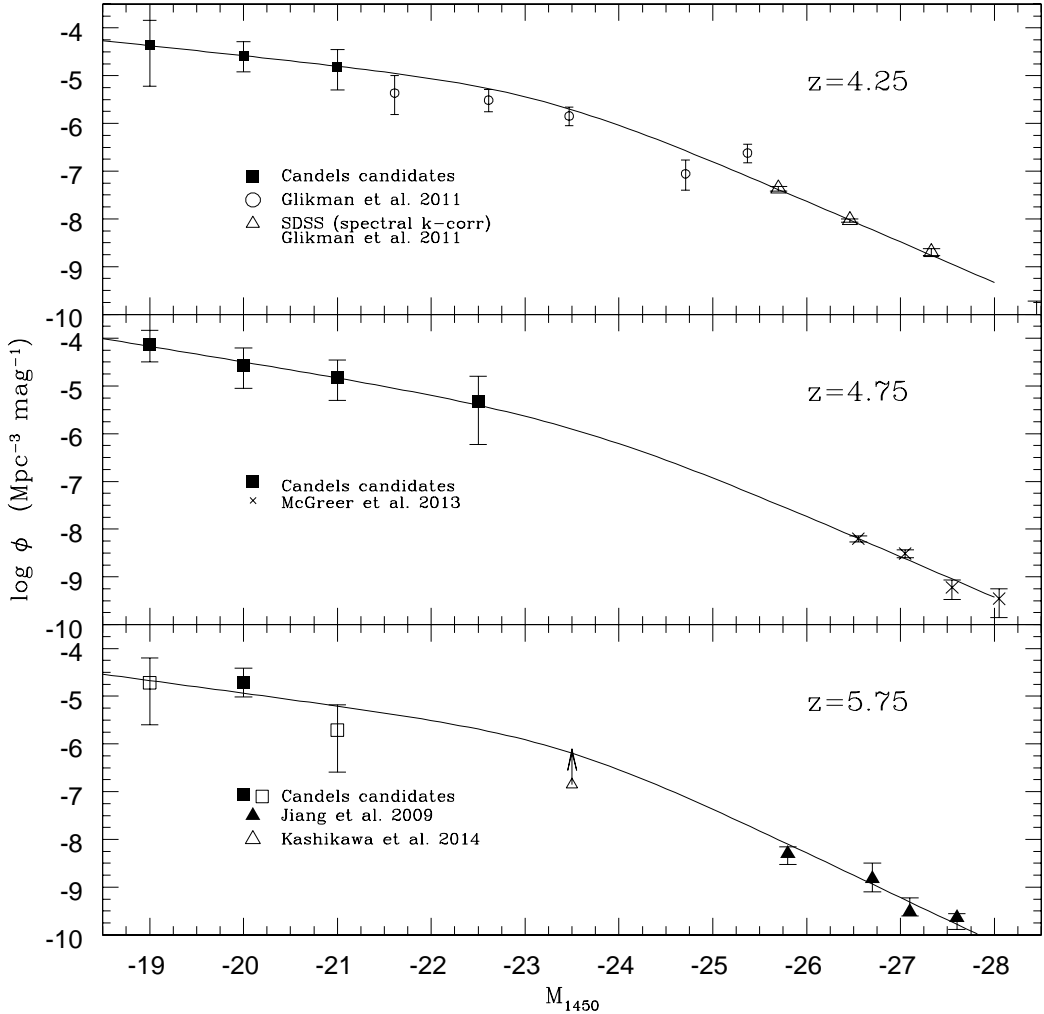}}
\caption{UV 1450 {\AA} AGN luminosity functions in various redshift bins. Different symbols represent different surveys as explained in the figure box.  Open squares in the highest redshift interval (bottom panel) represent LF bins derived from AGN candidates with more uncertain photometric redshifts (see Section  6.1)}
\end{figure*}

The resulting volume densities as a function of $M_{1450}, z$  have then been corrected for various sources of incompleteness. The first correction is due to the gradual H band bending of the galaxy counts which corresponds e.g. to a 50\% drop at $H=27$ in area \#4 of the present sample. In this respect, a correction factor has been included as a function of area and $H$ magnitude in the volume estimate of each object. The second correction is due to the fact that, for each H magnitude, only relatively bright X-ray AGN candidates (with $f_X>1.5\times 10^{-17}$ erg s$^{-1}$ cm$^{-2}$) can be detected in our $z\geq 4$ parent sample (see Figure 3).  This corresponds to the loss of AGN candidates with e.g. $F_X/(\lambda f_H)\lesssim 0.1$ at $H\sim 27$. The incompleteness fraction is derived from the same $X/H$ distribution observed above the X-ray flux threshold. The objects have been weighted considering that their H distribution in Figure 3 is affected by the faint slope of the UV rest-frame luminosity function and by the incompleteness in the H band counts. We adopted a faint end slope -1.5 and count incompleteness as derived in section 2. The two effects tend to compensate and the final correction is not sensitive to slope changes of $\sim 0.2$. Finally, correction factors to volume densities typically of the order of 10-20\%  have been applied to volume densities after taking into account spatial fluctuations in the X-ray flux limits for each position of the X-ray detection.

Given the very poor statistics we only evaluated the Poisson contribution to the errors in each LF bin adopting the recipe by \cite{gehrels86} valid also for small numbers. In this respect the errors shown in the figure represent lower limits respect to the true values. Indeed for example cosmic variance can play a significant role in the present deep pencil beam survey increasing the uncertainties in the derived volume densities.

The resulting luminosity functions are shown in figure 4 and table 3 for $M_{1450}\leq -18.5$ where the width and position of the bins have been selected to allow where possible a homogeneous statistics among the bins.
It is clear that our faint AGN candidates are sampling the faint end of the AGN UV luminosity function in the absolute magnitude range $-18.5\gtrsim M_{1450}\gtrsim -22.5$
and with densities at the faint end between $10^{-5}\lesssim \phi \lesssim 10^{-4}$ Mpc$^{-3}$ mag$^{-1}$. Thus the presence of even few faint high $z$ AGN candidates in the GOODS-S field changes appreciably the abundance of the AGN population at $z=4-6.5$ and its cosmological significance especially in the context of the cosmic reionization.

\begin{table*}
\caption{AGN luminosity functions, emissivities and photoionization rates$^1$}
\centering
\begin{tabular}{c c c c c c cc c c c r}
\hline\hline
$\Delta z$ & M$_{1450}$ & $\log \phi_{obs}$ & $\log \phi_{corr}$ & N$_{obj}$ & $\beta$ & $\gamma$ & M$_{break}$ & $ \log \phi^*$ & $ \epsilon_{1450}$ & $ \epsilon_{912}$ & $ \Gamma$ \\
\hline\hline
$4-4.5$ & & & & & 1.52 & 3.13 & $-23.2$ & $-5.2$ & 11.45 & 6.8 & 0.62 \\
 & $-19$ & $-4.9$ & $-4.3$ & 1 & & & & & & & \\
 & $-20$ & $-4.7$ & $-4.6$ & 3 & & & & & & & \\
 & $-21$ & $-4.9$ & $-4.9$ & 2 & & & & & & & \\

\hline
$4.5-5$ & & & & & 1.81 & 3.14 & $-23.6$ & $-5.7$ & 9.90 & 5.92 & $0.43$\\
 & $-19$ & $-4.5$ & $-4.1$ & 3 & & & & & & & \\
 & $-20$ & $-4.9$ & $-4.6$ & 2 & & & & & & & \\
 & $-21$ & $-4.9$ & $-4.8$ & 2 & & & & & & & \\
 & $-22.5$ & $-5.3$ & $-5.3$ & 1 & & & & & & & \\

\hline
$5-6.5$ & & & & & 1.66$^2$ & 3.35 & $-23.4^2$ & $-5.8$ & 4.2 & 2.5 & $0.12$\\
 & $-19$ & $-5.2$ & $-4.7$ & 1 & & & & & & & \\
 & $-20$ & $-5.1$ & $-4.7$ & 3 & & & & & & & \\
 & $-21$ & $-5.7$ & $-5.7$ & 1 & & & & & & & \\

\hline
\end{tabular}\\
$^1\phi$ in units Mpc$^{-3}$, $\epsilon$ in units of $10^{24}$ ergs s$^{-1}$ Hz$^{-1}$ Mpc$^{-3}$ , $\Gamma$ in units of $10^{-12}$ s$^{-1}$. $\phi_{corr}$ is the volume corrected for incompleteness in the X/H distribution. $^2$Value has been fixed to average in the two lower $z$ intervals.
\end{table*}

In fact to provide a first estimate of the total AGN emissivity at $z\geq 4$ a shape of the luminosity function has been derived in each redshift bin connecting the volume densities estimated from our sample with that of the brightest high $z$ SLOAN QSO sample where selection effects respect to the morphological appearance and X-ray properties are thought to be small. In other words, no  X-ray QSOs with strong absorption in the rest-frame optical/UV are expected at the brightest magnitudes where the Sloan sample should be representative of the overall AGN population.  We adopted a double power-law shape to connect the two samples in each redshift bin of the type
\begin{equation}
\phi=\frac{\phi^*}{10^{0.4(M_{break}-M)(\beta-1)}+10^{0.4(M_{break}-M)(\gamma-1)} }
\end{equation}

The best fit parameters are shown in table 3. Note that in the highest redshift bin the faint end slope and the break magnitude have been fixed to the average values in the two lower redshift  intervals because of the lower statistics. In general however, it is clear from figure 4 that the break magnitude and the related volume density are not well constrained from the present composite sample.  For this reason, the fitted volume density at the break is likely a lower limit and larger AGN surveys of intermediate depth are needed to fix the point.
In the lowest redshift bin $4\leq z < 4.5$ we have also shown for comparison the points of the luminosity function derived from a faint sample of type 1 QSOs with stellar appearance by \cite{glikman11}. It is to note how the X-ray detection criterion used in the present paper is able to reach higher volume densities including  AGNs with different morphological and spectral properties. In the highest redshift bin $5\leq z <6.5$ we have also shown in Figure 4 the recent point by \cite{kashikawa14} obtained from an optical spectroscopic survey where AGN candidates are selected from standard color selection, assuming a stellar appearance. We have added a vertical arrow to note that this value could represent a lower limit respect to combined NIR/X-ray surveys like that presented here. A decrease in volume density by a factor $\lesssim 2$  from $z\sim 4.25$ to $z\sim 5.75$ is present in our sample, smaller than found at higher luminosities in UV and X-ray selected surveys (e.g. \cite{vito14}).
In summary, our new data give a first estimate of the overall shape of the UV AGN luminosity function in the absolute magnitude interval  $-18.5\gtrsim M_{1450}\gtrsim -28$. A flattening of the LF is needed to match the volume density of the brightest QSOs derived from the Sloan survey to that of our faint AGN population. The global behaviour of the luminosity function is used in the next section to derive the high redshift evolution of the UV emissivity.

\section{AGN Hydrogen Ionizing Emissivity and Photoionization rate}

We now proceed to compute the contribution of our predicted luminosity functions to the AGN hydrogen ionizing emissivity. This is given 
by 
\begin{equation}\begin{split}
& \epsilon_{ion} (z)=\langle f\rangle \epsilon_{912}=\\
&\langle f\rangle \int \phi(L_{1450},z)L_{1450}\left({1200\over 1450}\right)^{0.44}\left({912\over 1200}\right)^{1.57}dL_{1450} 
\end{split}\end{equation}
where  $\langle f\rangle$ is the average escape fraction of ionizing radiation from the AGN host galaxies and $\epsilon _{912}$ is the emissivity produced by AGN activity. The shape of the AGN SED from $\lambda=1450$ {\AA} to $\lambda=912$ {\AA} has been represented by a double power law with slopes adopted after \cite{schirber03,telfer02} and \cite{vandenberk01}. 
In the integral we have considered all the AGNs a factor 100 brighter or fainter than the break luminosity (corresponding to a faint limit of $M_{1450}\sim -18$).
Considering sources even 100 times fainter would increase the emissivity only by a factor $\sim 10$\% given the rather flat slope of the luminosity function.
The uncertainties in the integrated emissivities are computed combining the uncertainties in the normalization and bright end slope for the highest redshift bin since the faint end slope and break have been fixed.
For the lower redshift bins the uncertainties are due to uncertainties on the normalization and faint end slope only (i.e. fixing the break luminosity and bright end slope). In this respect the widths of the error bars in emissivity and photoionization rates should be considered as lower limits.
It is to note that in the sampled redshift intervals the main contribution to the emissivity comes from AGNs near the predicted break magnitude  $M_{1450}\sim -23.5$.
It is just at this magnitude where volume densities shoub be estimated with greater accuracy but this requires surveys of comparable depth extended on larger areas.

In figure 5 and Table 3 the Lyman continuum ionizing emissivities  $ \epsilon_{24} $ in units of $10^{24}$ erg s$^{-1}$ Hz$^{-1}$ Mpc$^{-3}$ derived from the estimated AGN luminosity functions are shown as a function of redshift. We assume   $\langle f\rangle=1$  as a reference value for the ionizing escape fraction as observed in most bright AGNs.

The prediction of the model proposed by \cite{giallongo12} is also shown in figure 5 for comparison.  The model is based on the standard  $\Lambda$CDM scenario for galaxy formation and evolution already described in \cite{menci05} and includes a physical description of the Super Massive Black Hole growth at the center of galaxies as mainly due to the inflow of cold gas destabilized by minor merger events (\cite{cavaliere00,menci08}). The model also includes a self consistent description of the ionizing emissivity of AGNs based on the blast-wave feedback mechanism (\cite{cavaliere02,lapi05}). The blast-wave produced in the interstellar medium sweeps out the gas allowing ionizing UV photons free to escape from the AGN host galaxy into the IGM. 
The model prediction tends to overestimate the observed emissivities by factors two to five as the redshift increases from $z\sim 4$ to $z\sim 6.5$. In the same figure 5 the AGN model prediction by \cite{haardt12} is also shown for comparison. The latter model underestimates by a larger amount the observed values up to a factor 20 at $z\sim 6$.
Overall the new data are in between the two models, not far from the predictions by the \cite{giallongo12} model, and show a gradual decline of the ionized emissivity in the probed redshift interval $z=4-6.5$. To look whether this decline is consistent with the evolution of the IGM ionization level we computed the hydrogen photoionization rate predicted from our emissivity and compared it with that inferred from the analysis of the Lyman-$\alpha$ forest in high redshift QSO spectra.

The photoionization rate per hydrogen atom $\Gamma_{-12}$ in units of $10^{-12}$ s$^{-1}$ was computed as in \cite{giallongo12}  including the recent update on the mean free path (mfp) of ionizing photons in the IGM derived by \cite{worseck14}:
\begin{equation}
\Gamma_{-12}(z)\simeq 0.6 {\epsilon_{24}(z)\over 3+\mid \alpha_{UV}\mid} \left(\Delta l\over 65Mpc\right){\left(1+z\over 4.5\right)}^{3-\eta}
\end{equation}
where $ \alpha_{UV}=-1.57$ (\cite{schirber03}). The value $\alpha _{\nu}=-1.57$ adopted for $475<\lambda<912$ {\AA} is consistent with the recent evaluation with the COS spectrograph at HST by \cite{shull12} and \cite{stevans14} $\alpha_{\nu}=-1.41$. The power-law index $-\eta$ describes the decrease in redshift of the proper mean free path (mfp) $\Delta l$ of ionizing photons in the IGM due to the increase in redshift of the Lyman Limit absorption systems. We adopt $\eta=5.4$ and the normalization of the mfp to 65 Mpc at $z=3.5$ as found by \cite{worseck14} in the redshift range $z=4.4-5.5$. It is to note that the equation assumes that only ionizing sources within one absorption length contribute to the cosmic reionization. The values derived from our CANDELS GOODS-S sample of AGN candidates show in figure 6 a gradual decline from $z=4$ to $z=6.5$ although slightly steeper than that derived for the emissivity due to the increase of the average IGM absorption with increasing redshift. 

Different data points derived in a model-dependent way from the IGM statistics observed in the Lyman-$\alpha$ forest of high $z$ QSO spectra are also shown for comparison as in \cite{giallongo12}. Some values depend on the UV background adopted in numerical simulations to reproduce the mean flux decrement observed in the Lyman-$\alpha$ forest of QSO spectra (\cite{faucher08}, \cite{bolton05}, \cite{becker13}, \cite{fan06}, \cite{wyithe11}). Other values are derived from the analysis of the proximity effect at the highest redshifts $z\sim 5-6$ by \cite{calverley11}. Different methods produce the significant scatter shown in figure 6 despite the data are scaled to the same IGM temperature-density relation as in \cite{giallongo12}. This is due to uncertainties on modelling the gas density and temperature distributions in the IGM at very low densities (\cite{miralda00}) and within the proximity region surrounding each QSO (\cite{calverley11}). 

It is important to note from figure 6 that the photoionization rates at $z>4$ derived by our AGN luminosity functions are consistent with the photoionization levels of the IGM derived from the Lyman-$\alpha$ statistics in the same redshift range. 
This would imply a reionization occurring at redshifts not much larger than $z\sim 7$ since moderately bright ionizing sources with a maximum comoving density 
of $\sim 2\times 10^{-5}$ Mpc$^{-3}$ at $z\sim 6$ have a typical separation of $\sim 3$ proper Mpc, only a factor 2 smaller respect to the derived proper mean free path of ionizing photons in the IGM at $z\sim 6$ \cite{worseck14}. The latter scenario would also be consistent with the large decline of strong Lyman-$\alpha$ emitting galaxies
observed between $z\sim 6$ and $z\sim 7$ which would suggest a large neutral hydrogen fraction in the IGM and a patchy reionization process already at $z\sim 7$
(\cite{pentericci11,treu13,pentericci14,tilvi14}).

\section{Discussion}

In the context of this preliminary analysis we should discuss some caveats which could significantly reduce our expectations about the contribution to the reionization by high redshift AGNs of intermediate-low luminosities. First of all some redshift estimates for our candidates could be wrong, despite the average high level of accuracy, reducing the number of high z AGNs. The second issue concerns the assumption about the predominance of the AGN UV spectral energy distribution even in cases where the optical spectra appear affected by significant stellar contribution or look very steep. A further issue concerns the assumption of a high escape fraction of ionizing photons for the global AGN population, including the X-ray absorbed AGN fraction. Finally we discuss the implication that an ionizing AGN population at very high redshift would have for an early HeII reionization of the IGM.

\subsection{Redshift reliability}
Concerning the reliability of the derived redshift distribution in our sample we should note that the evaluation of the photometric redshifts becomes progressively more uncertain for fainter sources with featureless SEDs. For this reason we have shown in the Appendix  the probability redshift distributions PDF(z) for our candidates.
In this context, as already stated, the estimate of photometric redshifts for sources at $z>4$ mainly relies on the statistical significance of the Lyman $\alpha$ forest ($<1216$ {\AA}) and Lyman break at 912 A rest frame wavelength. In fact, the expected escaping Lyman continuum emission from the sources, even assuming  $\langle f\rangle\simeq 1$,  would be strongly depressed by IGM absorption causing a flux dropout. As a consequence, the redshift uncertainty at $z=5-6$  is related to the flux dropouts near the Lyman-$\alpha$ and Lyman edge almost independently of the assumed galaxy or AGN spectral library. Our assumption is corroborated by the good agreement we found between the 5 available spectroscopic redshifts and the photometric estimates of the CANDELS catalog. It is clear however that in cases where the spectrum is particularly steep the evidence for the presence of any Lyman break weakens, increasing consequently the uncertainty in the redshift estimate. This is particularly true for some of the $z>5$  objects. Indeed we have already excluded from the LF analysis the object 29323 because of its too peculiar SED and PDF(z). Thus five sources have been used for the LF estimate in the highest redshift bin, among them 20765, 28476 and 33160 have the most uncertain redshift estimate as shown in Figure A.1. To get a rough estimate of the uncertainties involved in the derived volume densities of faint AGNs at $z>5$ we have repeated the estimate of the luminosity function at the highest redshift bin excluding these sources. Two of them are the only sources in the faintest and brightest LF bins ($M_{1450}=-19$ and $M_{1450}=-21$) that would be removed. We have indicated with different symbols these uncertain LF bins. Two sources remain in the LF bin at $M_{1450}=-20$ slightly decreasing the average volume densities by $-0.1$ to $\log \phi =-4.8$. This value would still be consistent with the two-power law extrapolation adopted to estimate the UV emissivity.  Finally we note that some of these sources are relatively bright at 8 micron but this by no means represents a prior against a high redshift solution for the estimated redshift. Indeed one of the best studied X-ray absorbed faint AGN in our catalog, 273 (e.g. \cite{vanzella08}), which has also been observed by ALMA (\cite{gilli13}), has a robust spectroscopic redshift at $z=4.76$ and an 8 micron apparent AB magnitude of 21.5. Also the source 14800 at spectroscopic redshift $z=4.82$ shows a relatively bright IRAC continuum at levels of 22.5.

 Very recently \cite{weigel15} have searched for $z>5$ AGNs in the same GOODS-S field using almost the same CANDELS dataset finding no convincing AGN candidates. The different result depends on their adopted procedure. They have looked for $z>5$ sources starting from the \cite{xue11} X-ray selected catalog, looking for plausible optical drop-outs and/or photometric redshifts in the CANDELS images. Our sample is based on NIR selection in the H band and reaches fainter X-ray fluxes respect to the \cite{xue11} catalog. Moreover photometric redshifts in our sample have been obtained  from the CANDELS multiwavelength catalog derived from the B to the 8 micron SPITZER band that included careful SPITZER deblended photometry. As an example one source in our sample, (9713 not included in the \cite{xue11} )  has a spectroscopic confirmation at $z=5.7$. 
In our sample of 22 AGN candidates only 2 are in common with the \cite{xue11} catalog with an estimated redshift  $z>5$. One of the two, 29323 (X156), has already been excluded from our analysis as stated above. The second object, 33160 (X85), appears at $z\sim 6$ although with a larger uncertainty and we have already checked in the previous paragraph the effects of its removal from the sample. 


\subsection{UV spectral shape}
With few broad band filters often showing only upper limits it is impossible to disentangle the relative AGN and stellar contributions to the SED of our faint $H=24-27$ mag sources. This is why we do not include in our sample AGN candidates at $z<4$. Indeed only at very high redshifts ($z=4-6$) the photometric redshifts are mainly constrained by strong  Lyman absorptions by the intergalactic medium intervening along the line of sight, almost independently of the intrinsic spectral shape assumed (AGN or galaxy template).
If the sources are at very high redshifts, the observed optical bands ($<1$ micron) give access to the rest frame UV SED shortward of 3000 {\AA}.
The presence of X-ray detection in our sample at levels $\log L_X\gtrsim 43$ (erg s$^{-1}$) implies probable AGN emission. As a consequence, somewhere in the UV part of the spectrum the AGN emission should emerge above the stellar spectrum. We assume that this happens at wavelengths just shortward of 1500 {\AA}. In particular we assume that our 1450 {\AA} luminosity is mainly due to AGN emission. Of course to verify this hypothesis a detailed spectroscopic information would be required which is beyond the capability of the present instrumentation due to the faintness of our candidates. Nevertheless this assumption is supported by the distribution of the rest frame $X/UV$ flux ratio in our sample (which at z>4 corresponds to the $X/H$ ratio distribution shown in Figure 3).
Our AGN candidates are confined  in the X/H interval between about 0.01 and 0.3-0.5  (in the same units of Figure 3) usually occupied by the AGN population in brighter AGN surveys as shown e.g. in \cite{fiore12} for the COSMOS high redshift AGNs (\cite{civano11}). Also the few SLOAN QSOs at $z=6$  show $X/H$ fluxes which occupy the same region up to $0.2-0.3$ although most of them have ratios below 0.1 because they are very bright optically selected sources (and so with a significant  X-ray weak fraction).
If the rest frame UV fluxes were dominated by stellar emission this would imply an intrinsic AGN X/H flux ratios much (e.g. >10 times) greater than generally observed in brighter high redshift AGNs. Of course both rest-frame X-ray and UV fluxes could be produced by stellar populations but this seems unlike for X-ray luminosities $\log L_X\gtrsim 43$ (erg s$^{-1}$). By the way the suggestion provided by the SED models about the possible presence of Balmer breaks appears not compelling and not necessarily caused by stars. Rest-frame optical nebular emission lines (at $z=5-6$) could mimick Balmer breaks whose contribution to broad band filters increases with (1 + z) as noted by  \cite{shaerer09}.

Finally to check if the average spectral shape assumed to estimate the ionizing emissivity from the UV AGN luminosity function is appropriate for our sample,
we have derived a first rough guess of the average power-law slope in the UV rest-frame interval 1300-1700 {\AA}, just around 1500 {\AA}. The average slopes have been derived from e.g. the V,I,z bands at $z\sim 4$ and Y,J bands at $z\sim 6$. The average/median slope $\alpha_{\nu}$ was found -1.4/-1.2 (excluding object 9323) although the dispersion is significant. In our paper we have adopted a flatter slope from 1450 down to 1200 {\AA} (-0.44) and a steeper one shortward of 1200 {\AA} (-1.57). Assuming the derived average slope -1.4 (instead of -0.44) just shortward of 1450 {\AA} reduces the photoionization rates by about 10 percent only. To halve the derived photoionization rates in Fig.6 we should adopt a much steeper -2.5 slope for all the AGNs in the luminosity function. The average UV slope derived in our sample,-1.4, appears thus consistent with that (-1.57) adopted in the paper for the overall AGN population from $M_{1450}=-18$ to $M_{1450}=-28$. 

In summary, although we can not reliably quantify the stellar contribution to the optical/UV emission we assume that in our X-ray sources the SED shortward of the rest-frame 1450  {\AA} is dominated by AGN emission. This assumption appears consistent with the typical AGN X/H (X/UV rest-frame) flux ratio found in our sample, which would imply an average UV-X spectral slope similar to that found in AGNs.

\subsection{Ionizing escape fraction}
One of the main uncertainties concerns the assumption about the escape fraction of UV photons from AGNs. In fact, we assume that all ionizing photons are able to escape and ionize the surrounding IGM. This is usually almost true for bright type 1 QSOs where the optical depth is $\tau_{LL}<1$ at the Lyman limit position corresponding to the systemic redshift of the QSOs. Any appreciable absorption is present at shorter wavelengths due to the intervening IGM along the line-of-sight (e.g. \cite{sargent89}).  Our faint sample of AGN candidates lacks spectroscopic information except for the 5 cases shown in table 2 and could include a significant fraction of absorbed AGNs. Hints of excess absorption have been derived in the X-ray spectral analysis of high redshift X-ray selected AGNs but the reliability of any measure is hampered by the fact that the photoelectric cutoff of the aborption goes out of the spectral region sampled by the Chandra observations already at $z>3$ (see e.g. \cite{vito13}). Moreover to allow a first estimate of the hydrogen column density with few photon counts, a fixed slope (typically $\Gamma=1.8$) of the intrinsic X-ray power-law is often adopted and  a solar metallicity is always assumed despite observational hints of supersolar metallicities in the black hole neighbourhood. In fact flatter $\Gamma$ values (\cite{vito13}) and supersolar metallicities could both significantly reduce the estimated hydrogen column density. Some statistics on the relative fraction of obscured/unobscured QSOs of intermediate X-ray luminosity ($43\lesssim \log L_X\lesssim 44$) 
shows that the fraction of obscured AGNs at high redshifts could be $\sim 50$\% (\cite{hasinger08,merloni14}) although there is no tight correlation between    X-ray absorption and the presence of broad emission lines since some X-ray absorbed AGNs show broad emission lines in their rest-frame UV spectra, as outlined by \cite{fiore12}. Moreover recent COS spectral analysis of FUV spectra of low $z$ AGNs of different luminosities ($\log L_{UV}=44-46.5$) and spectral types (Sy 1, Sy 1.2, Sy 1.5, QSOs) by \cite{shull12} and \cite{stevans14} show no continuum edges at 912 {\AA}.   In this complex situation what we are assuming is that the presence of any AGN activity in high $z$, rest-frame UV selected galaxies - activity certified by X-ray emission at levels $\log L_X\gtrsim 43$ - is able to provide the needed mechanical and radiative feedback mechanisms to allow ionizing UV photons free to escape outside the host galaxies into the surrounding IGM with average escape fractions in the sample not much smaller than 1 (e.g. in the range  $\langle f\rangle\sim 0.7-1$).  It is to note that a reduction of the expected photoionization rate by e.g. 30\% respect to that shown in Figure 6, due to an assumed average escape fraction $\langle f\rangle\sim 0.7$, would be compensated by the re-emission of ionizing photons by the IGM due to radiative recombination (\cite{haardt96}) that in a framework of reionization by the AGN population could contribute by a factor $\sim 1.25$ to the cosmological photoionization rate at $z\gtrsim 5$ (Haardt, private communication). 

Feedback at various intensity levels is indeed emerging from the most recent observations as an ubiquitous activity in  AGNs, independently of their spectral type (e.g. \cite{cicone14} and reference therein).   Due to the feedback mechanisms in the AGN host galaxies some escaping ionizing photons could also come from young stars in the hosts. For a given AGN/stellar flux ratio at 1500 {\AA},  the stellar ionizing spectrum at $\lambda \lesssim 912$ {\AA} is expected to be softer than shown by the AGN due to intrinsic Lyman continuum absorption in the stellar atmospheres. 
The latter absorption is essentially unknown and expected values depend on the spectral synthesis models adopted, on the age and metallicity assumed for the stellar populations in star forming galaxies. Typical values for the $F_{1500}/F_{912}$ flux ratio, which is a measure of the intrinsic stellar absorption, are $\sim 3$ for normal star forming galaxies. However for our AGN host galaxies at $z>4$ the age and metallicity of the associated starburst activity could be  $\lesssim 10^7$ yr and $\sim 0.2 Z_{\odot}$, respectively,  implying intrinsic $F_{1500}/F_{912}$ ratios $\sim 2$ or smaller. These values could be not far from the ratio 1.7 derived from our adopted AGN spectral shape. In other words the ionizing UV spectra from a very young population in AGN host galaxies could be almost as hard as that of the AGNs at wavelengths just shortward of the Lyman edge at 912 {\AA}.  
Of course with decreasing wavelengths shortward of 912 {\AA} the AGN and stellar spectral shapes are progressively more and more different and at the HeII edge (228 {\AA}) only the AGN SEDs can contribute to the HeII reionization.

\subsection{HeII reionization}
For this reason, in our proposed scenario  for the cosmic reionization, the hardness of the ionizing UV spectrum provided by the AGN population would produce an early significant reionization of the HeII content in the IGM. In fact, there is some consensus that HeII reionization could be completed later in the universe, at $z\sim 3$, respect to hydrogen, supporting a scenario where galaxies with much softer UV spectra dominate the ionization of hydrogen at very high redshifts until AGNs become so numerous at $z\sim 3$ to provide the required hard photons to reionize HeII. The issue has been already discussed in \cite{giallongo12}  in the framework of an AGN dominated UV background where the HeIII volume filling factor was computed as a function of decreasing redshift under various assumptions. In particular it was recognized that the presence of optically thick HeII clouds between AGNs reduce the mean free path of HeII ionizing photons, according to \cite{bolton09}, allowing a gradual increase of the HeII reionization as the redshift decreases up to a full reionization at $z\sim 3$. Of course a detailed evolutionary history of the hydrogen and helium reionization requires 3D hydrodynamical models however the simplified model suggests that a modest degree of inhomogeneity in the HeII distribution could delay its reionization history. 
In this context, recent direct observations of HeII spectral regions in 17 bright QSOs at $z\sim 3$ with HST/COS by \cite{worseck14b} seem to suggest a mild evolution of the HeII reionization with increasing redshift allowing the possibility to get the bulk of the HeII reionization in the universe at $z>4$.  On the other hand recent detailed hydrodynamical simulations of the IGM (\cite{compostella14,puchwein14}) enligthen the need for an increase of the hydrogen and HeII photoionization rates respect to what predicted by the \cite{haardt12} model at very high redshifts. These analyses if confirmed could support a reionization scenario driven by AGNs.

\begin{figure}
\centering
\scalebox{1.2}[1.2]{\hspace{-0.5cm}\includegraphics[width=\hsize]{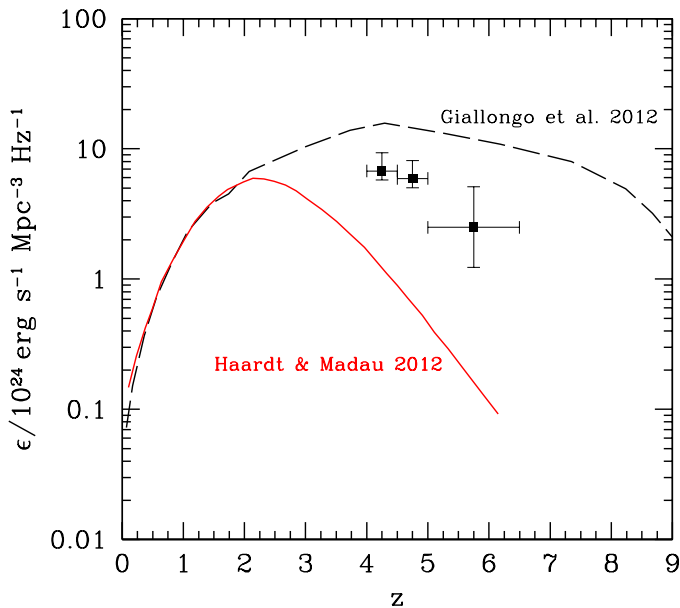}}
\caption{Cosmic ionizing emissivity by AGNs as a function of redshift assuming $\langle f\rangle=1$. Black squares are from our sample. Red continuous curve is from the Haardt and Madau (2012) model. The long dashed curve is from the Giallongo et al. (2012) model.}
\end{figure}

\begin{figure}
\centering
\scalebox{1.2}[1.2]{\hspace{-0.5cm}\includegraphics[width=\hsize]{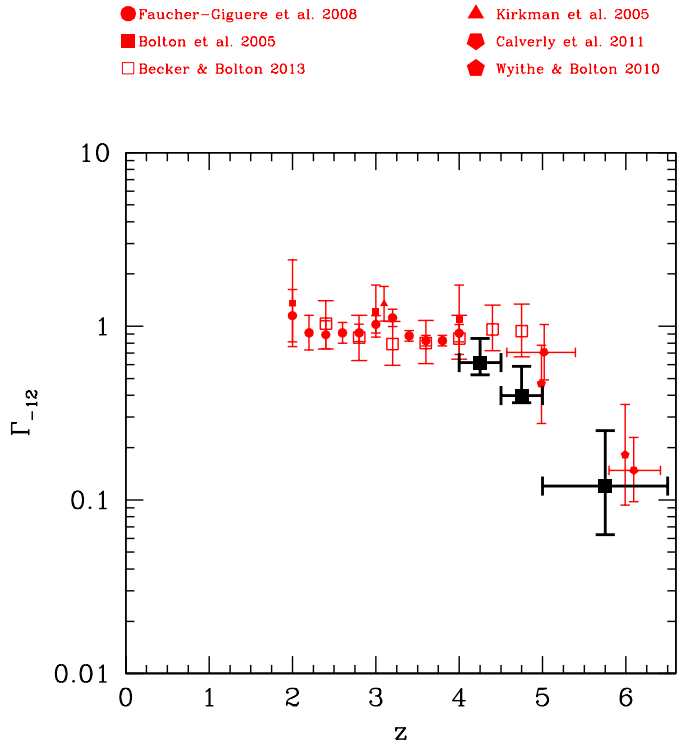}}
\caption{Cosmic ionizing photoionization rate $\Gamma_{-12}$ in units of $10^{-12}$ s$^{-1}$ produced by AGNs as a function of redshift assuming $\langle f\rangle=1$. Black filled squares represent the predicted contribution by faint AGNs from the GOODS-S sample. Other small red symbols are the values inferred from the ionization status of the IGM as derived from the Lyman $\alpha$ forest analysis in high $z$ QSO spectra.}
\end{figure}

\section{Summary}

Thanks to the completion of deep multiwavelength surveys it is now possible to apply specific selection procedures to reveal the very high redshift AGN population with intrinsic luminosities more typical of local Seyfert galaxies rather than bright Sloan QSOs.

More specifically we have selected AGN candidates at $z>4$ in the CANDELS GOODS-S field starting from the NIR H band selection of the galaxy parent sample, which corresponds to UV rest-frame selection, down to a very faint magnitude limit $H\leq 27$.  We have used photometric and few spectroscopic redshifts derived from the CANDELS catalog and we have finally extracted from these high $z$ galaxies the subsample of AGN candidates  showing X-ray detection above
$F_X\sim 1.5\times 10^{-17}$ erg cm$^{-2}$ s$^{-1}$ (0.5-2 keV), correponding to a probability of spurious detection of $2\times 10^{-4}$ in the Chandra 4Msec GOODS-S field.

We have made a preliminary estimate of the selection function aimed at deriving a correction for the sample incompleteness. We have derived average volume densities $10^{-5}\lesssim \phi \lesssim 10^{-4}$ Mpc$^{-3}$ mag$^{-1}$ in the magnitude interval $-18.5\gtrsim M_{1450}\gtrsim -22.5$. Thus the presence of even few faint high $z$ AGN candidates in the GOODS-S field increases appreciably the cosmological significance of the AGN population at $z=4-6.5$ especially in the context of the cosmic reionization.

To make a first estimate of the ionizing UV emissivity of the AGN population at these high $z$ we have derived for the first time the shape of the UV luminosity function from $M_{1450}\sim -18.5$ to $M_{1450}\sim -28$ adding the bright QSO samples from the Sloan survey which is less affected by significant incompleteness. The UV luminosity function is well represented by a double power-law although the position of the break is highly uncertain. The values obtained from a simple fit are used to derive an ionizing emissivity of the order of $\epsilon_{24} \sim 6.5$ at $z=4-5$ and $\epsilon_{24}\sim 2.5$ at $z\sim 6$, assuming an average AGN UV spectral shape and an escape fraction of ionizing UV photons $\langle f\rangle=1$.

The derived emissivity has then been used to estimate the hydrogen photoionization rate in the same redshift interval which is of the order of $\Gamma_{-12}\sim 0.1$ at $z\sim 6$, consistent with the value required to keep highly ionized the intergalactic medium observed in the Lyman-$\alpha$ forest of bright QSO spectra at the same redshift. This scenario points toward a late reionization epoch at redshifts not much higher than $z=7$ as suggested by the rapid decrease in the number of Lyman alpha emitters and by the recent lower thompson optical depth suggested by recent preliminary Planck data\footnote{\url{http://www.cosmos.esa.int/web/planck/ferrara2014}}.

We have discussed some caveats which could reduce the volume density and/or ionizing emissivity of the high redshift AGN population described in the present work. More specifically the large uncertainties involved in photometric redshifts for sources with steep, almost featureless, spectral energy distribution could spuriously enhance the total number of $z>5$ AGN candidates in the GOODS-S field. However, after careful evaluation of this bias, we have shown that possible changes in the AGN volume densities provide values still consistent with the adopted shape of the UV luminosity function in the redshift interval $5<z<6.5$. 	A further important caveat is related to the escape fraction of ionizing photons from the AGN host galaxies which should be on average $\langle f\rangle \gtrsim 0.5$ to ensure enough ionizing photons in the universe. At present it is not clear if fainter AGNs ionize their neighborhood as well as the brighter QSO population does.

We finally emphasize that, given the large uncertainty in the present data, the aim of this preliminary study is not focused on deriving a reliable shape of the AGN UV luminosity function but is rather devoted to establish if faint AGNs can be definitely ruled out as main contributors to the cosmic reionization, as generally thought. The main outcome of the present preliminary analysis goes in the opposite direction. Given the large uncertainties involved there is room for a significant contribution to reionization by the $z>4$ AGN population under reasonable physical assumptions. Looking for faint NIR selected high-$z$ galaxies with even marginal X-ray detection in the deepest available NIR and X-ray images can be a promising technique to reveal a large number of Lyman Continuum emitting sources. Applying the same selection strategy and spectroscopic follow up to other fields with similar characteristics like e.g. GOODS-N and COSMOS will provide a composite larger sample and consequently a more reliable estimate of the AGN UV luminosity function and photoionization rates up to redshift  $z\gtrsim 6$ where the cosmic hydrogen reionization is still in action.

\begin{acknowledgements}
We acknowledge useful comments by the referee which have improved the analysis in the discussion section. We thank F. Haardt for discussion and for giving us preliminary estimates of the IGM ionizing contribution at high redshifts. We thank V. Khaire for noticing a typo in an earlier version of table 3. We acknowledge financial contribution from the agreement ASI-INAF I/009/10/0. This work is based on observations taken by the CANDELS Multi-Cycle Treasury Program with the NASA/ESA HST, which is operated by the Association of Universities for Research in Astronomy, Inc., under NASA contract NAS5-26555. Observations were also carried out using the Very Large Telescope at the ESO Paranal Observatory under Programme IDs LP186.A-0898, LP181.A-0717,
LP168.A-0485, ID 170.A-0788, ID 181.A-0485, ID 283.A-5052 and the ESO Science Archive under Programme IDs 60.A-9284, 67.A-0249, 71.A-0584, 73.A-0564, 68.A-0563, 69.A-0539, 70.A-0048, 64.O-0643, 66.A-0572, 68.A-0544, 164.O-0561, 163.N-0210, 85.A-0961 and 60.A-9120. This work is based in part on observations made with the Spitzer Space Telescope, which is operated by the Jet Propulsion Laboratory, California Institute of Technology under a contract with NASA. Support for this work was provided by NASA through an award issued by JPL/Caltech.
AF acknowledges the contribution of the EC FP7 SPACE project ASTRODEEP (Ref.No: 312725).
\end{acknowledgements}

\Online

\begin{appendix}

\section{Spectral energy distributions and redshift probability distribution functions}

In this section we show the spectral energy distributions (SEDs) and the probability distribution functions (PDFs) of the photometric redshift estimates for each AGN candidate derived from the \cite{dahlen13} analysis. Most of the candidates in figure A1 show PDFs confined at $z>4$ with only small wings at $z<4$. The most uncertain redshifts being for the objects 20765, 28476, 29323, 33160 although the larger PDFs distributions appear mostly distributed at $z>4$. It is to note that the latter three objects at $z>6$ are among the most uncertain due to combination of  power-law shape of their SEDs and  faintness of the sources. Note also that for some of these objects, the bayesian photometric redshift does not correspond to the peak of the PDF but it is the result of a weighted average around the maximum.
\begin{figure}
\centering
\scalebox{1}[1]{\hspace{-0.5cm}\includegraphics[width=\hsize]{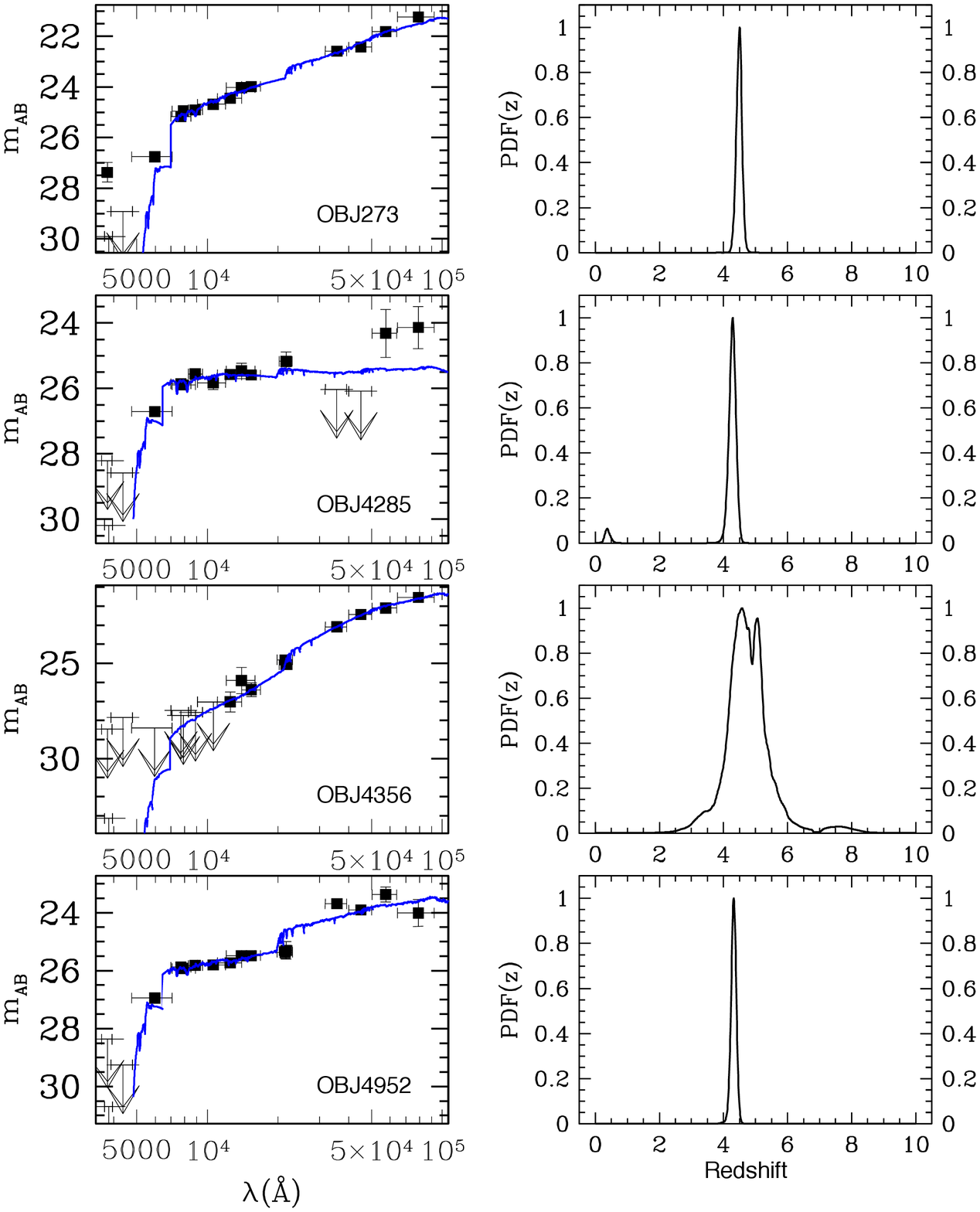}}
\end{figure}

\begin{figure}
\centering
\scalebox{1}[1]{\hspace{-0.5cm}\includegraphics[width=\hsize]{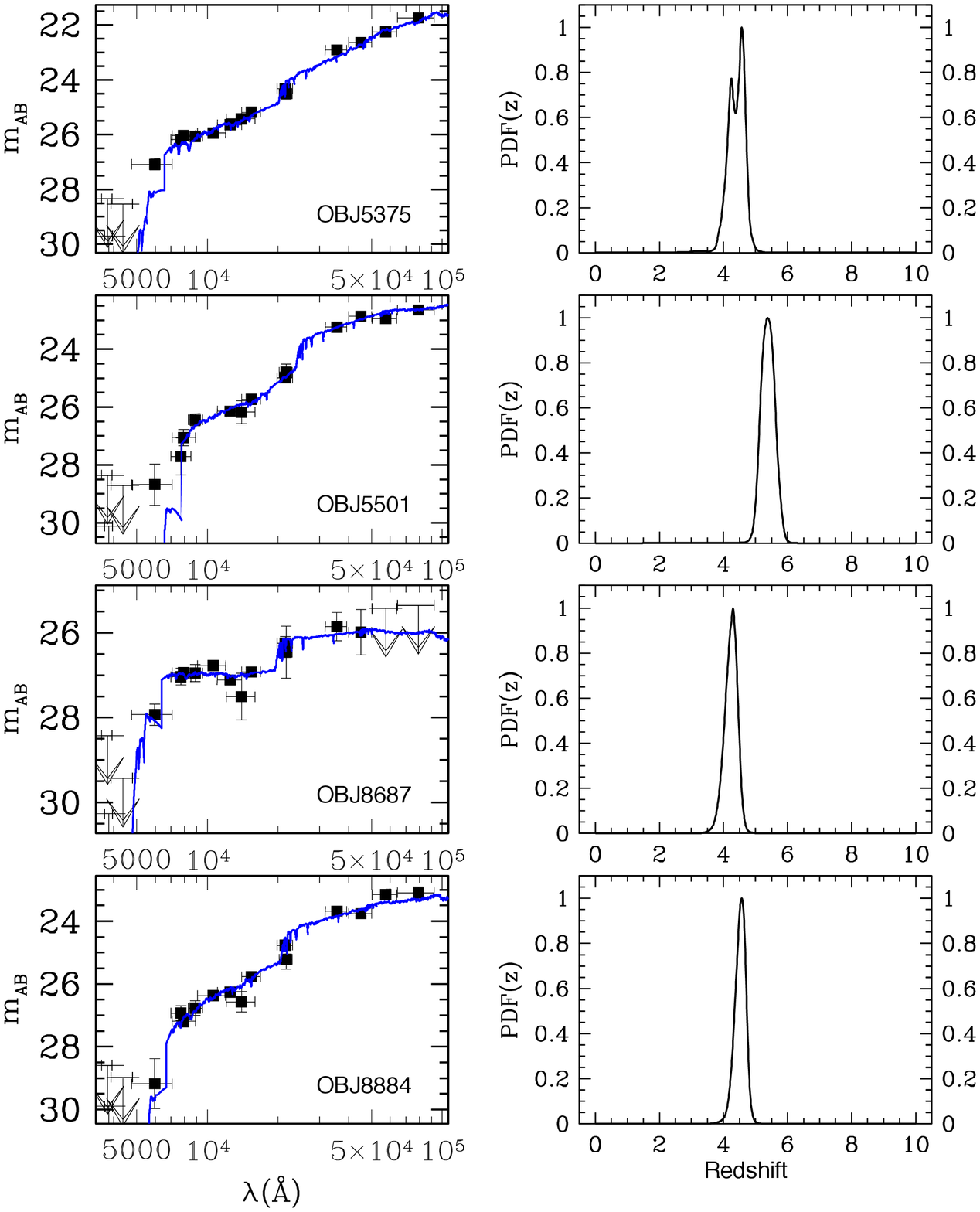}}
\end{figure}

\begin{figure}
\centering
\scalebox{1}[1]{\hspace{-0.5cm}\includegraphics[width=\hsize]{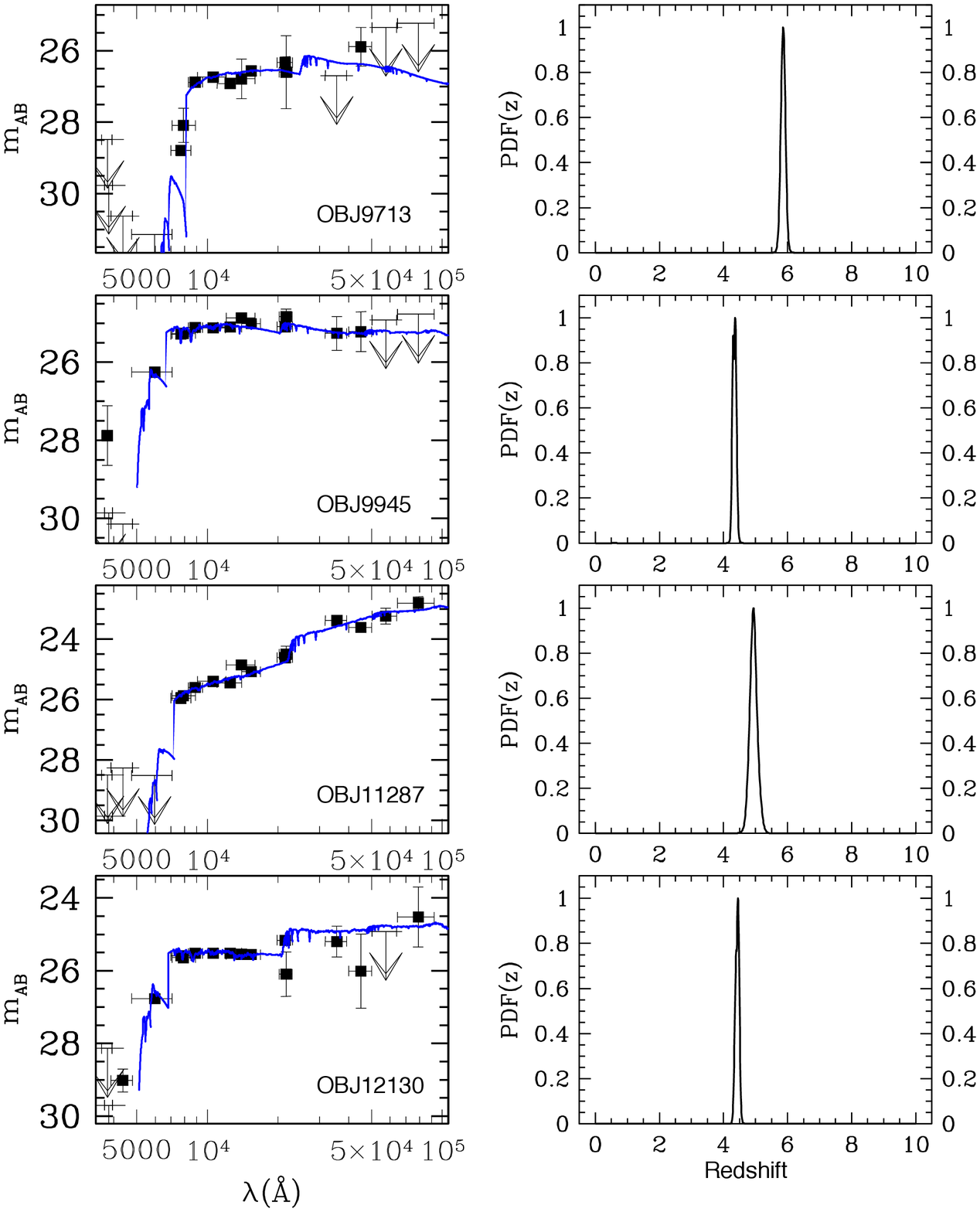}}
\end{figure}

\begin{figure}
\centering
\scalebox{1}[1]{\hspace{-0.5cm}\includegraphics[width=\hsize]{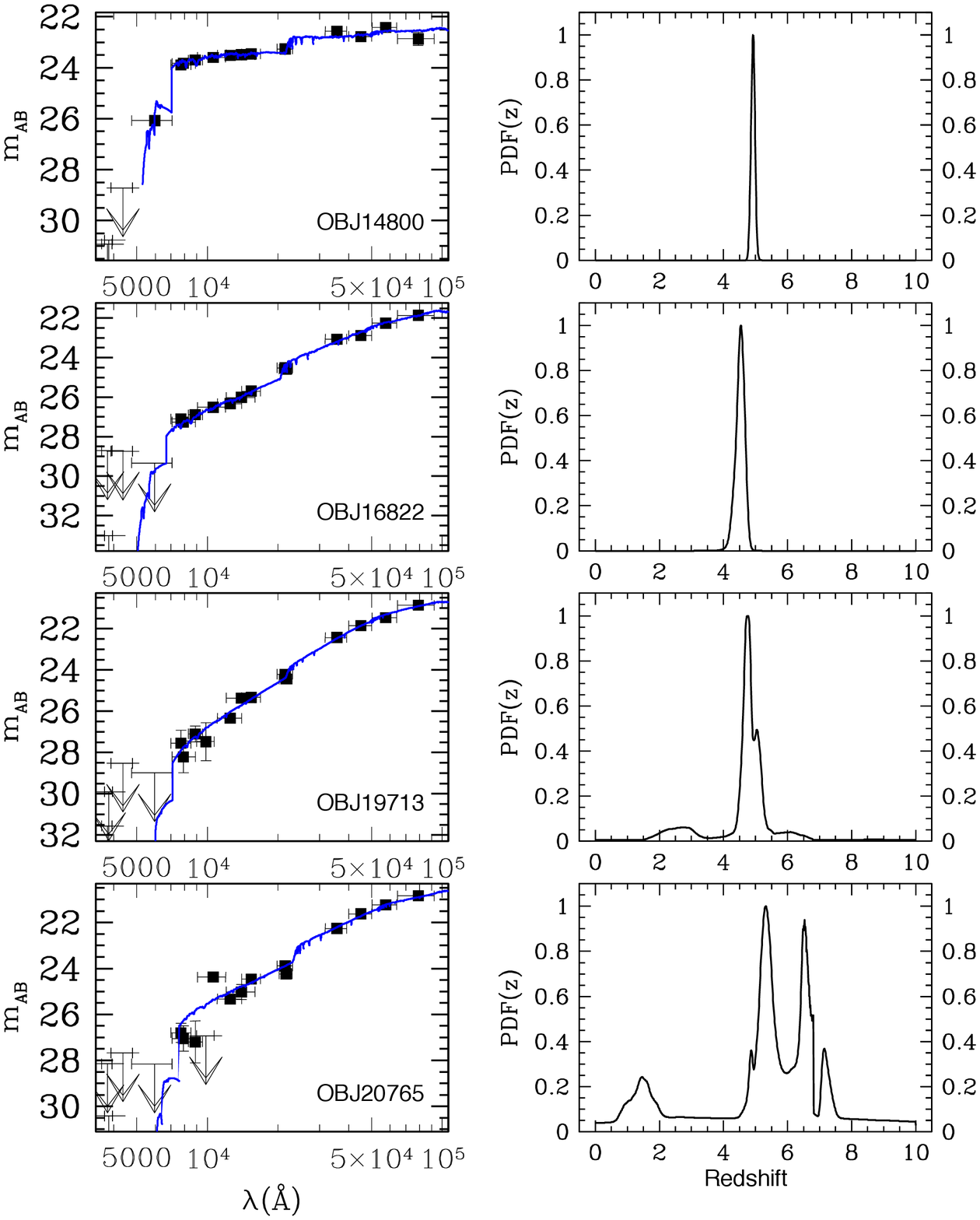}}
\end{figure}

\begin{figure}
\centering
\scalebox{1}[1]{\hspace{-0.5cm}\includegraphics[width=\hsize]{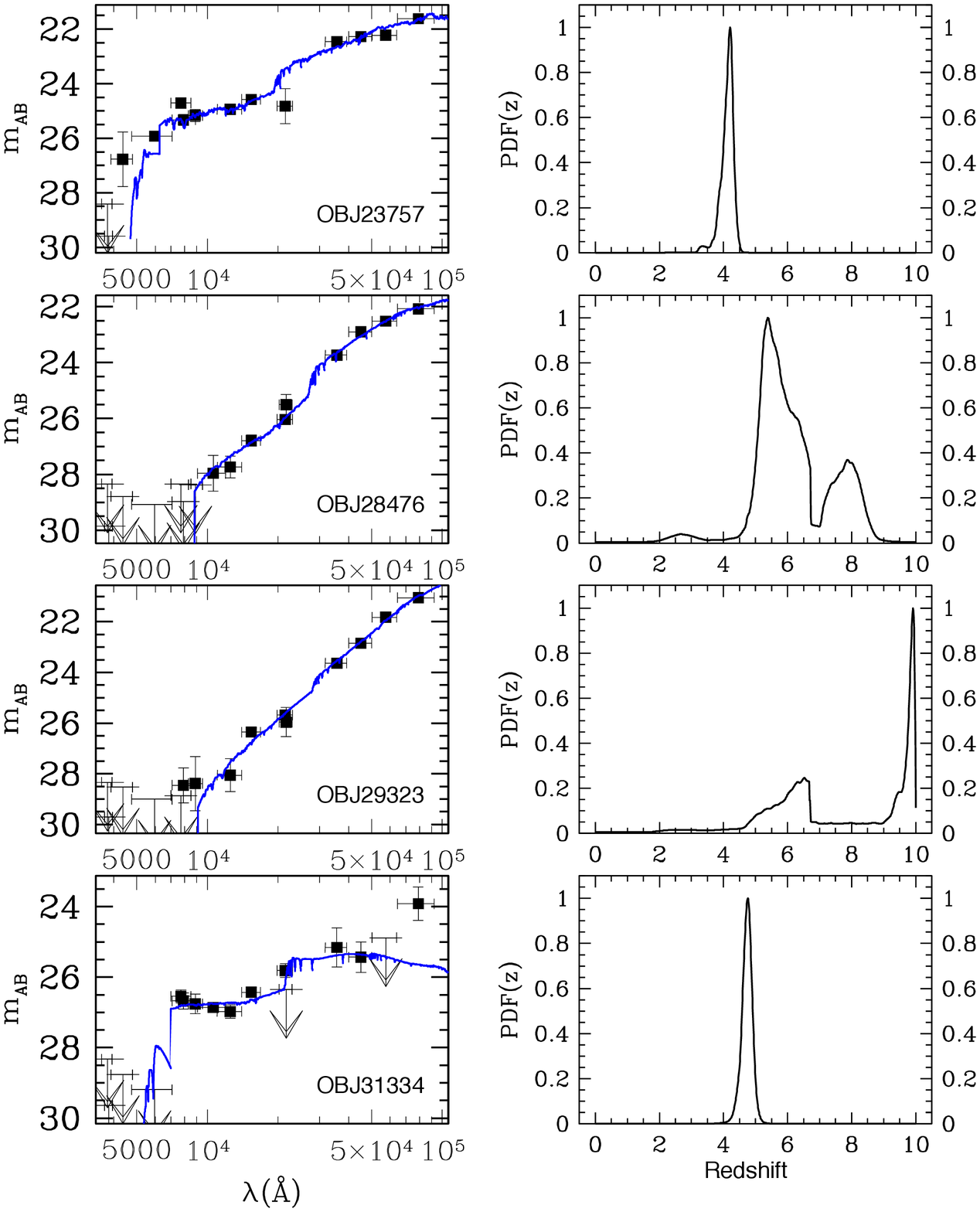}}
\end{figure}

\begin{figure}
\centering
\scalebox{1}[1]{\hspace{-0.5cm}\includegraphics[width=\hsize]{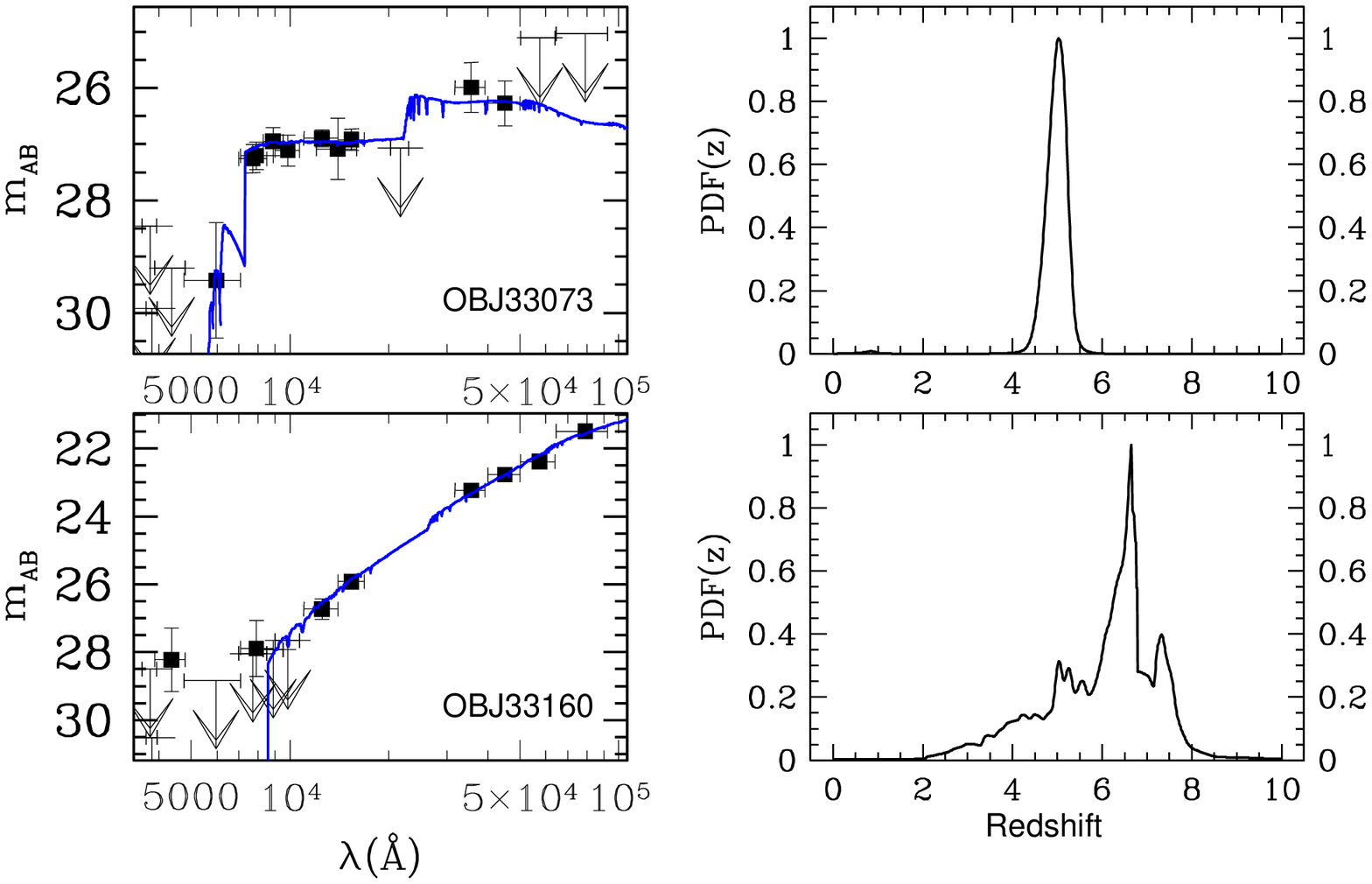}}
\caption{Spectral energy distributions and probability distribution functions of photometric redshifts normalized to 1 at the peak for all the AGN candidates shown in Table 2. 1-$\sigma$ upper limits are also shown as downward arrows.}
\end{figure}

\section{X-ray position accuracy}

In this section we show an example of the relative position accuracy between the X-ray and H band image. We have selected the object 28476 already shown in Figure 1 which appears located near two brighter H band sources. In Figure B.1 we also show the position accuracy of a  bright source on the left bottom corner which results comparable to that of 28476. In general the position accuracy between the H and X-ray sources is $\lesssim 1$ arcsec for all the AGN candidates of our sample. They have off-axis Chandra positions $<9$ arcmin.

\begin{figure}
\centering
\scalebox{1}[1]{\hspace{-0.5cm}\includegraphics[width=\hsize]{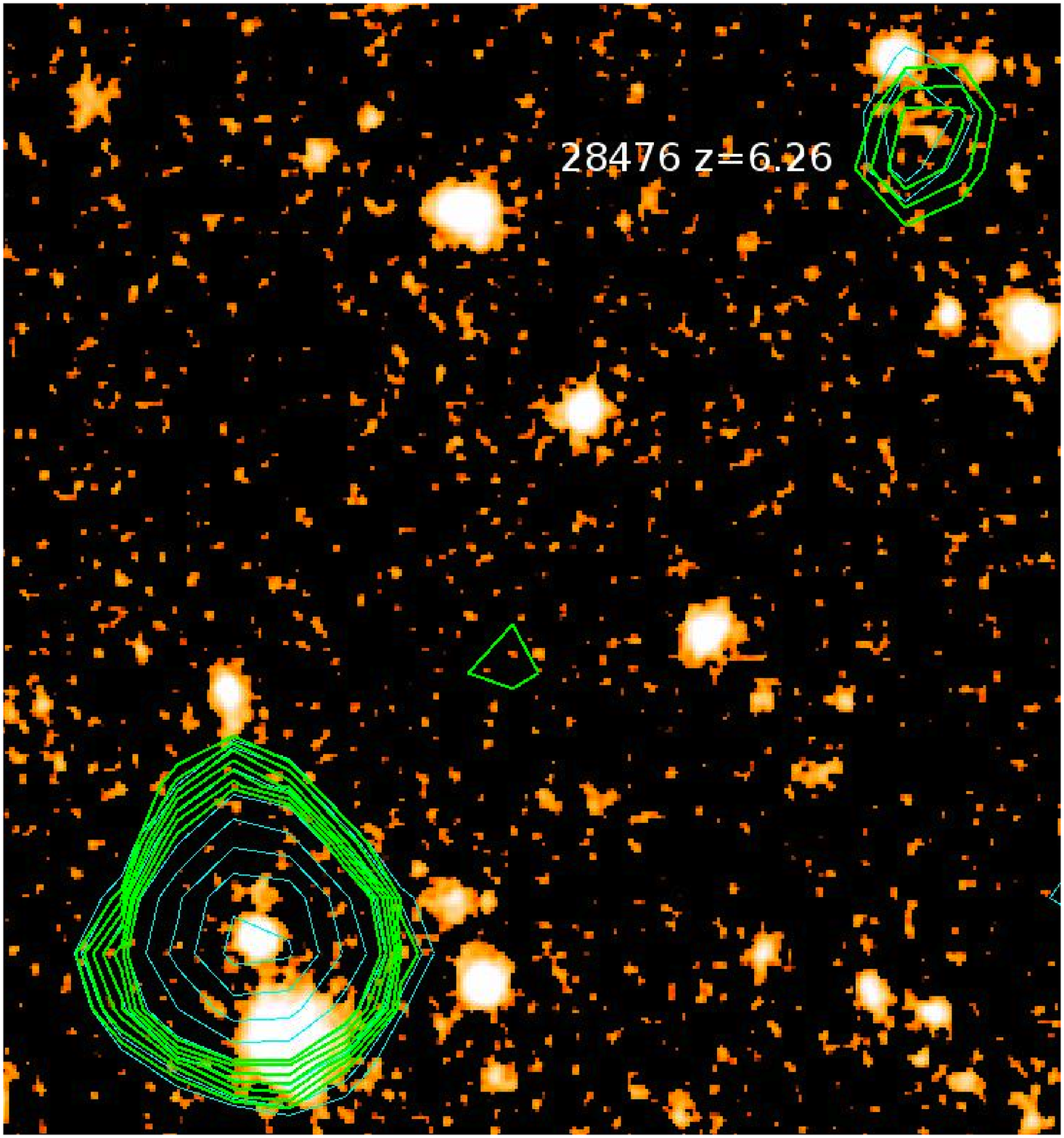}}
\caption{H band image around the AGN candidate 28476 (upper right) shown in Figure 1. The size is 30 arcsec. Cyan and green contours are X-ray detections in the 0.8-4 keV and 0.5-2 keV bands, respectively. Position accuracies are within 1 arcsec. }
\end{figure}

\end{appendix}

\end{document}